\documentclass{aa}

\usepackage{newtxtext,newtxmath}
\usepackage{amsmath,ulem}
\usepackage{physics}
\usepackage{multirow}
\usepackage{caption}
\usepackage{float}

\usepackage[T1]{fontenc}
\DeclareRobustCommand{\VAN}[3]{#2}
\let\VANthebibliography\thebibliography
\def\thebibliography{\DeclareRobustCommand{\VAN}[3]{##3}\VANthebibliography}

\usepackage{dblfloatfix}
\usepackage{graphicx}	
\usepackage{amsmath}	
\usepackage{amssymb}	
\usepackage{hyperref}
\usepackage{xcolor}
\usepackage{float}

\title{Impact of the disk magnetization on MHD disk wind signature}

\author{
Sudeb Ranjan Datta \inst{1},
Susmita Chakravorty\inst{2}, Jonathan Ferreira\inst{3}, Pierre-Olivier Petrucci\inst{3}, Timothy R Kallman \inst{4}, Jonatan Jacquemin-Ide \inst{5}, Nathan Zimniak \inst{3}, Joern Wilms \inst{6}, Stefano Bianchi \inst{7}, Maxime Parra \inst{3,7}, Ma\"ica Clavel \inst{3}
}

\institute{
Astronomical Institute of the Czech Academy of Sciences, Bo\v{c}n\'\i -II 1401, Praha 4, Prague, 141~00, Czech Republic\\
\email{datta@asu.cas.cz}
\and
NE 211, IISc, Bengaluru, India, 560012 \\
\email{write2susmita@gmail.com}
\and
Univ. Grenoble Alpes, CNRS, IPAG, 38000 Grenoble, France
\and
NASA Goddard Space Flight Center, Greenbelt, MD 20771, USA
\and
Northwestern University, CIERA Evanston, IL 60201
\and
Friedrich-Alexander-Universität Erlangen-Nürnberg, Erlangen, Germany
\and
Dipartimento di Matematica e Fisica, Università degli Studi Roma Tre, via della Vasca Navale 84, 00146 Roma, Italy\\
}

\titlerunning{effect of magnetization on absorption}  
\authorrunning{Datta et al.}

\date{Received September 15, 1996; accepted March 16, 1997}
\begin{document}


\abstract
{Observation of blue-shifted X-ray absorption lines indicates the presence of wind from the accretion disk in X-ray binaries. Magnetohydrodynamic (MHD) driving is one of the possible wind launching mechanisms. Recent theoretical development makes magnetic accretion-ejection self-similar solutions much more generalized, and wind can be launched even at much lower magnetization compared to equipartition value, which was the only possibility beforehand.}
{Here, we model the transmitted spectra through MHD driven photoionized wind - models which have different values of 
 magnetizations. We investigate the possibility of detecting absorption lines by the upcoming instruments XRISM and Athena. Attempts are made to find the robustness of the method of fitting asymmetric line profiles by multiple Gaussians.}
{We use photoionization code XSTAR to simulate the transmitted model spectra. To cover the extensive range of velocity and density of the wind spanned over a large distance ($\sim10^5$ gravitational radii), we divide the wind into slabs following a logarithmic radial grid. Fake observed spectra are finally produced by convolving model spectra with instruments' responses. Since the line asymmetries are apparent in the convolved spectra as well, this can be used as an observable diagnostic to fit for, in future XRISM and Athena spectra. We demonstrate some amount of rigor in assessing the equivalent widths of the major absorption lines, including the Fe XXVI Ly$\alpha$ doublets which can be clearly distinguished in the superior quality, future high resolution spectra.}
{Disk magnetization becomes another crucial MHD variable that can significantly alter the absorption line profiles. Low magnetization pure MHD outflow models are dense enough to be observed by the existing or upcoming instruments. Thus these models become simpler alternatives to MHD-thermal models. Fitting with multiple Gaussians is a promising method to handle asymmetric line profiles, as well as the Fe XXVI Ly$\alpha$ doublets.
}
{}


\maketitle

\section{Introduction}

Observations of X-ray binaries (XRBs) frequently reveal blue-shifted absorption lines, which are the signatures of outflowing wind from the accretion disk (\citealt{Lee2002, Miller2004, Miller2006, Miller2008, Neilsen2009, Ueda2009, Kallman2009, King2012, Miller2016}, for review see \citealt{Trigo2016, Ponti2016}). Over the last two decades, XMM-Newton, Chandra (and currently NICER, NuSTAR) have been providing a wealth of data on winds in X-rays. X-ray winds are mainly observed in highly inclined XRBs during the high-soft state of their outburst (\citealt{Ponti2012, Parra2023}). The dependence on the inclination suggests an equatorial geometry. Absence of wind signatures in X-ray spectra, in the canonical hard states, cannot be attributed to mere changes in the illuminating spectra (\citealt{Lee2002, Ueda2009, Ueda2010, Neilsen2011, Neilsen2012}) and there is still no consensus about the reason for this dependence on spectral states. Thermodynamic instability could be one possible reason (\citealt{Chakravorty2013, Bianchi2017, Petrucci2021}). Recently wind signatures have been observed in the hard states as well, but in infrared and optical (\citealt{Rahoui2014,Darias2019,Ibarra2019,Cuneo2020,Sanchez2022,Darias2022}). This suggests the presence of winds during the entire outburst, albeit with state dependant changes in the physical properties of the wind. Wind signatures also change from one source to another, and in some for the same source from one observation to another, even in very similar spectral states (e.g. \citealt{Neilsen2012, Parra2023}) possibly due to an inhomogeneous medium along the line of sight (LOS).

Matter from the accretion disk around compact objects can be launched through different mechanisms: magnetohydrodynamic (MHD) (\citealt{Blandford1982, Ferreira1993a, Ferreira1995, Contopoulos1994, Contopoulos1995, Ferreira1997, Miller2006A}), thermal (\citealt{Begelman1983, Woods1996, Higginbottom2017, Tomaru2020, Tetarenko2020}), and radiation (\citealt{Icke1980, Shlosman1993, Proga2002, Higginbottom2023}) driving or even a combination (e.g magneto-thermal \citealt{Casse2000b}, see also \citealt{Waters2018}). For XRBs, the matter is highly ionized compared to the winds from AGN which diminishes the line driving possibility and makes the radiation driving inefficient. Both MHD and thermal mechanisms are promising sources for wind driving in XRBs. 

MHD driving at equipartition magnetic field strength is like a bead (matter lifted from the accretion disk) on a wire (the magnetic field line) anchored in the disk and co-rotating with it. If the magnetic field line is sufficiently inclined, the bead is ejected out due to centrifugal force, overcoming the gravitational attraction (\citealt{Blandford1982}). Vertical gradient of magnetic pressure can also play a dominant role in launching winds depending on the coupling between different components of the magnetic field. Thermal driving, on the other hand depends on heating the disk material. The scale height of the accretion disk increases with its radius, which makes the outer region inflated and irradiated by the radiation from the central part of the disk. This heats up the material, and ejection is achieved when thermal velocity overcomes the escape velocity (\citealt{Begelman1983}).

Over the last two decades, MHD models developed by two groups, namely a) \cite{Ferreira1993a} and b) \cite{Contopoulos1994}, have been used on several occasions to bridge the gap between theoretical solutions and wind observations in XRBs  \citep{Chakravorty2016, Chakravorty2023, Fukumura2010, Fukumura2017, Fukumura2021, Fukumura2022}. The primary difference between these two classes of MHD models lies in the connection of the winds with the disk. Accretion-ejection solutions of \citet{Ferreira1997, Casse2000b, Jacquemin2019} linked the density of the wind to the disk through the accretion rate, while wind solutions of \citet{Blandford1982, Contopoulos1995} treat the density of the wind as a free parameter, independent of the disk. Based on the solutions of \cite{Contopoulos1994, Contopoulos1995}, \cite{Fukumura2017} showed that MHD-driven winds can explain the observation of blue-shifted absorption lines in black hole systems across different mass scales.

\begin{figure*}[b]
\centering\includegraphics[width=\textwidth]{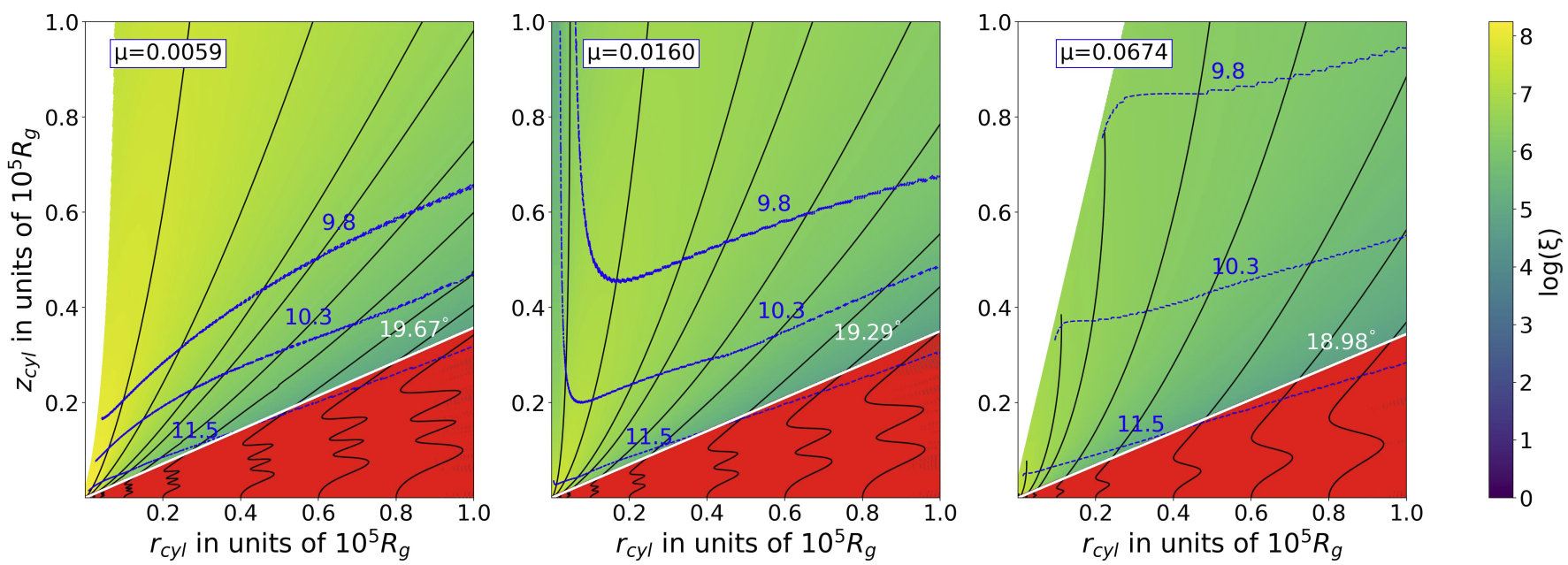}
\caption{Variation of $\xi$ in the r-z plane for different magnetizations, $\mu=0.0059, 0.0160, 0.0674$ (from left to right). The Compton thick region with N$_h>10^{24.18}$ along the line of sight is shown in red. The white line shows the lowest possible Compton thin line of sight and the value of corresponding $i$ is written next to it. Black solid lines spread over the whole region are the poloidal magnetic field lines anchored at different radii. Density contours are shown in blue dashed lines. Corresponding log of density (in cc) is written just next to each density iso-contour.}
\label{fig_Nh_xi_distribution}
\end{figure*}

Modeling of absorption through a photoionized MHD wind is studied in detail by \cite{Chakravorty2016} (hereafter Paper I) where different classes (`cold' and `warm' solutions introduced as CHM and WHM in section \ref{sec_MHD_solns}) of MHD wind solutions are incorporated. The `cold' solutions (\citealt{Ferreira1997}) are purely MHD winds whereas `warm' solutions (\citealt{Casse2000b, Ferreira2004}) are magneto-thermal as they include external heating at the disk's surface which aid the MHD forces in lifting outflow material. `Cold' solutions turned out to be too tenuous and thereby inefficient to reproduce the observed wind signatures, whereas the denser `warm' solutions can do so. A typical 100 ks observation using the upcoming instruments onboard Athena or XRISM can easily detect the wind signatures of `warm' solutions (\citealt{Chakravorty2023}, hereafter Paper II) and also the line asymmetries, which is one of the prominent features of MHD driven winds (\citealt{Fukumura2022}). However, these `cold' and `warm' solutions were highly-magnetized, with a magnetic field energy near equipartition (with thermal pressure). In accretion disks in XRBs it is expected that the outer region which contributes mostly in absorption (\citealt{Chakravorty2013}, Paper I) will have much lower magnetization compared to inner region (\citealt{Ferreira2006, Petrucci2008, Marcel2018}). That is why recently developed generalized `cold' MHD solutions (\citealt{Jacquemin2019}) become crucial in providing denser (compared to high-magnetized `cold' solutions) wind at lower magnetization. One can expect that even without the need of any additional disk surface heating, these cold, low magnetized (CLM) solutions possibly will be able to provide the observed equivalent width of absorption lines. In addition, similarly dense solutions are possible at different magnetizations which further motivates us to study the effect of magnetization on the transmitted spectra. Thus, this paper is dedicated to find out the effect of disk magnetization in the transmitted spectra keeping other parameters nearly fixed; based on the solutions developed by \cite{Jacquemin2019}. We discuss the different classes of accretion-ejection solutions in more detail in section \ref{sec_MHD_solns}.

The plan of the paper is as follows. In section \ref{sec_MHD_solns}, we discuss the theoretical solutions and how different physical constraints are used to select the detectable  wind region. The spectrum radiated by the central region of the disk and incident on the wind is constructed in section \ref{sec_in_spectra}. The mechanism of XSTAR computation is detailed in section \ref{sec_xstar}. Section \ref{sec_results} describes the Results - understanding the XSTAR simulated `model spectra' in section \ref{sec_variation_with_mag}; and analysing the XRISM and Athena-like fakespectra in section \ref{sec_simulated_spectra}. Discussion (Section \ref{discn}) and Conclusion (Section \ref{sec_conclusions}) follow. Further details on MHD solutions, incident spectra and XSTAR computation are given in Appendix \ref{appendix_MHD_solns}, \ref{sed_details}, \ref{appendix_xstar_comp} respectively.

\section{Magnetohydrodynamic solutions}
\label{sec_MHD_solns}
\subsection{Overview of available solutions}
\label{sec_different_mhd_solns}
MHD ejection of matter can occur self-consistently along with accretion to the central accretor (\citealt{Ferreira1993a}). In fact, ejection of matter through the magnetic field line anchored in the disk helps significantly in transporting angular momentum outward and consequently helps in accretion. The basic assumption for these semi-analytical solutions is self-similarity in distance from the black hole \citep{Ferreira1993a}.

One major model quantity is the disk ejection efficiency $p$, defined as the exponent of the disk accretion rate $\dot M_a(r) \propto r^p$. This exponent\footnote{While in the theoretical papers this exponent is labeled $\xi$, we use here $p$ to not confuse with the ionisation parameter} cannot be larger than unity if ejection is to be powered by accretion alone \citep{Ferreira1997}. It controls the amount of mass lifted from the disk, hence the wind density. The maximum speed achieved along a magnetic surface anchored on a radius $r_o$ is $u_{max}= V_{Ko}\sqrt{2\lambda -3}$, where $V_{Ko}$ is the Keplerian speed at $r_o$ and $\lambda$ the magnetic lever arm \citep{Blandford1982}. It turns out that $\lambda\simeq 1 + 1/2p$ \citep{Ferreira1997} so that, depending on the value of the disk ejection efficiency $p$, an accretion-ejection solution can be used to model a jet or a wind. 

A second very important parameter is the disk magnetization $\mu$, defined at the equatorial plane as
\begin{equation}
    \mu= \frac{V_{Az}^2}{C_s^2} = \frac{B_z^2/\mu_0}{P_{tot}}
\end{equation}
where $V_{Az}$ is the vertical component of the Alfven velocity, $C_s$ is the sound speed, $B_z$ is the vertical component of the large scale magnetic field, $\mu_0$ is the magnetic permeability of vacuum, and, $P_{tot}=P_{gas}+P_{rad}$ is the sum of the kinetic and radiation pressures. Note that this definition does not include the turbulent magnetic field, only the laminar large scale vertical component present in the disk.

As time progresses, different classes of these solutions have been obtained and are outlined one by one in the following. 

\subsubsection{Cold, highly magnetized (CHM) solutions} 
It is natural to think that MHD effects will be best able to impact accretion-ejection structures when the magnetic field is near equipartition, with $\mu$ between $\sim 0.1$ and 1  \citep{Ferreira1995}. For these cold solutions, namely either isothermal or adiabatic outflows emitted from near-Keplerian thin disks, the typical disk ejection efficiency is found to be around $p\sim 0.01$. Paper I studied the physical properties of these CHM solutions (for different possible ejection indices) and concluded that they are not dense enough, regardless of their radial extent, to produce the observed winds.

\subsubsection{Warm, highly magnetized (WHM) solutions} 
One way to have denser winds at same magnetizations is to consider that the surface layers of the disk are heated up by irradiation or MHD turbulence, leading naturally to magneto-thermal models \citep{Casse2000b, Ferreira2004}. These WHM solutions can produce sufficiently dense observable winds (Paper I). It has also been tested that Athena and XRISM will be able to detect absorption lines from warm MHD outflows and even the line asymmetries should be detected with standard 100 ks observation (Paper II). 

While heating of the outer disk surface is actually expected due to irradiation by light coming from the central regions of the disk, these models suffer from the caveat that the amount of extra heating is a free parameter. 

\begin{figure*}
\centering\includegraphics[width=0.9\textwidth]{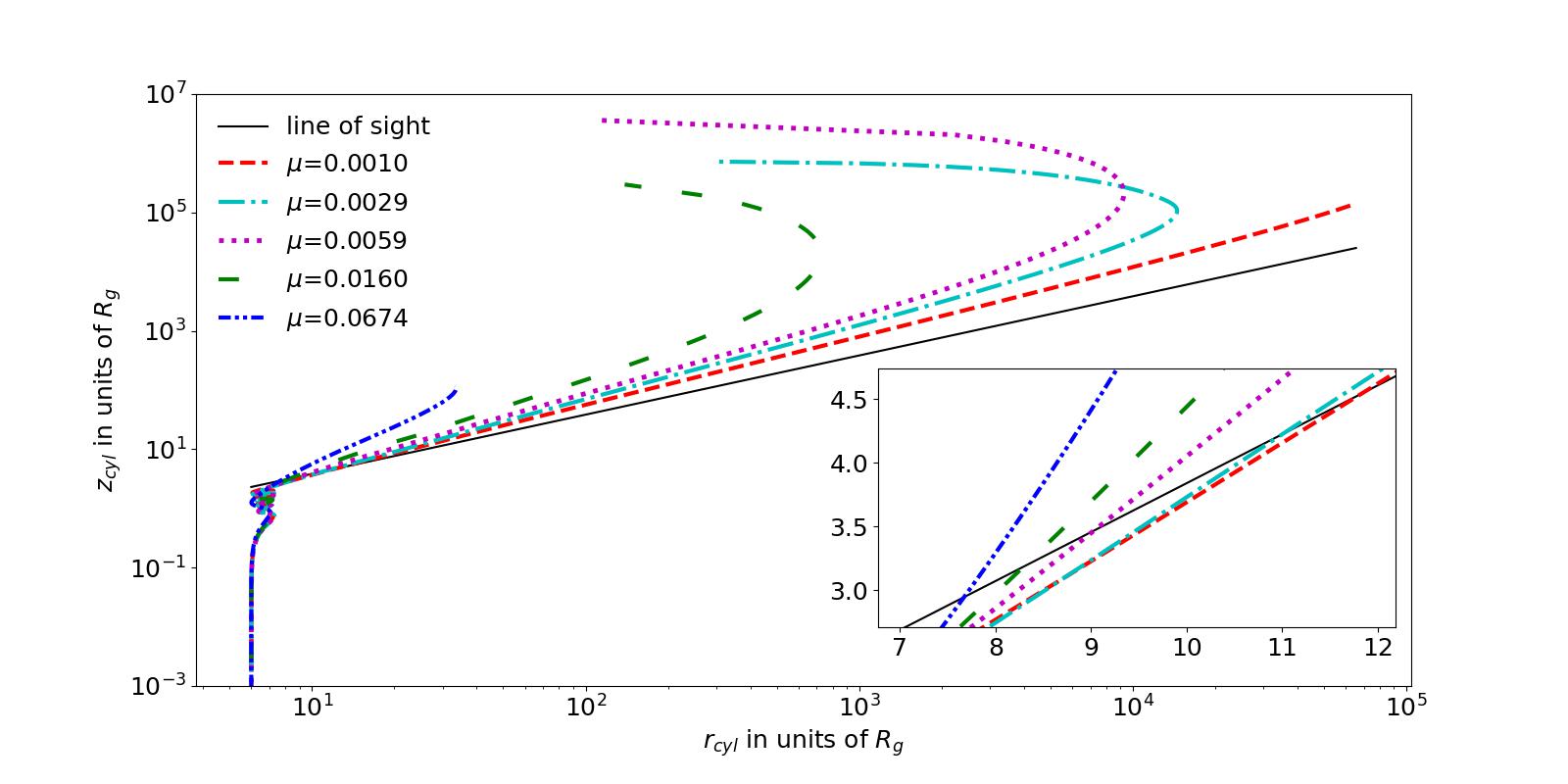}
\caption{Poloidal magnetic field lines for different magnetizations $\mu$. For each solution we show only one field line anchored at $6 R_G$ ($R_G$ is the gravitational radius). The solid black line indicates the line of sight at an inclination of i =  21$^\circ$ ($\theta = 69^\circ$). Inset shows a zoomed in view of the LOS crossing the magnetic field lines with different magnetizations. We see that the same LOS cuts the different field lines at different distances, due to the different geometry of the magnetic field lines. Please note that the axes of the inset are non-logarithmic.} 
\label{fig_mag_field_geometry}
\end{figure*}

\subsubsection{CLM (cold, low magnetized) solutions} 
Progressing more on the theoretical side of magnetized accretion-ejection solutions, \cite{Jacquemin2019} showed for the first time that significantly denser MHD wind solutions (as dense as WHM solutions) could be achieved with a much weaker magnetic field ($\mu$ ranging from $\sim 10^{-4}$ to 0.1.), without any requirement of an additional heating. At a much smaller magnetic field, ejection resembles a magnetic tower \citep{Ferreira1997, Lynden2003}, with a strong toroidal component uplifting the disk material.    
These CLM solutions are generalized in the sense that they also reproduce the CHM solutions when the strength of the field reaches equipartition (see Figs. 1 and 7 in \citealt{Jacquemin2019}). 

\begin{figure*}[b]
\centering\includegraphics[width=0.9\textwidth]{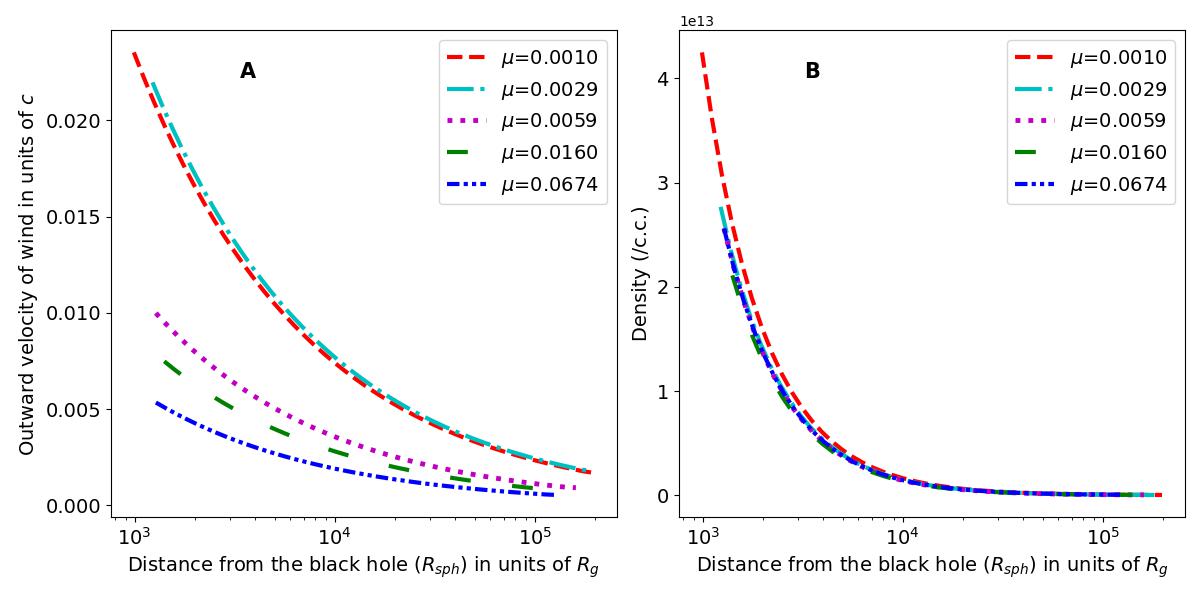}
\caption{Radial profile of (A) velocity and (B) density of wind along a line of sight of $i=21^\circ$ for different disk magnetizations $\mu$. Please note that although wind is present from a distance of $\sim$ 10 $R_g$ from the black hole (inset of Fig. \ref{fig_mag_field_geometry}), here we only show the part of the wind which has low enough ionisation to contribute significantly to absorption. That is why the radial distance from the black hole extends outwards from $\sim 500 R_g$ in this figure.}
\label{fig_vel_den_pro}
\end{figure*}

In the present paper, we focus on these new generalized CLM solutions. Earlier works already showed that although we can ignore the effect of the disk aspect ratio, $\epsilon$ (Paper I), the ejection index $p$ plays a very dominant role in determining the final transmitted spectra through the wind density (Paper II). With the discovery of the new generalized CLM solutions, a much larger range of magnetization become accessible to produce the wind. Thus, different dynamics (within the flows) can be achieved with different magnetization, even if the ejection index remains near constant for those different MHD solutions. In this paper we test the effect of disk magnetization on MHD disk winds. Hence we use MHD models for whom the ejection index is fixed to (nearly) the same value $p \sim 0.1$, but their magnetization $\mu$ vary through a range of about 1.8 dex.

\subsection{Effect of the disk magnetization on the wind structure}
\label{sec_hydro}

We choose five accretion-ejection CLM solutions with different disk magnetizations $\mu= 1.0\,\times\,10^{-3}, 2.9\,\times\,10^{-3}, 5.9\,\times\,10^{-3}, 16\,\times\,10^{-3}$ and $67.4\,\times\,10^{-3}$. All solutions have the same disk aspect ratio $\epsilon=H/R= 0.1$, the same MHD turbulence parameters and nearly the same disk ejection efficiency $p \sim 0.1$ (see Appendix \ref{appendix_MHD_solns}). The outflows are computed assuming isothermal magnetic surfaces, so the winds are `cold' (thermal effects are negligible) and are all of the ``magnetic tower'' type (see Fig~7 in \citealt{Jacquemin2019}).   

Despite the weak magnetic field, the dominant torque leading to accretion remains due to the wind (see Fig 9 in \citealt{Jacquemin2019}). This is because MHD turbulence due to the magneto-rotational instability scales with $\mu^{1/2}$ and is also small (see e.g. \citealt{Salvesen2016} and references therein). Nevertheless, accretion is subsonic, while the MHD wind carries away a significant portion of the released power (around 50\%). Figure~\ref{fig_Nh_xi_distribution} shows poloidal cuts of three such solutions: magnetic surfaces in black solid lines and background color showing the ionisation $\xi$\footnote{$\xi=L_{ion}/(nR^2_{sph})$, where $L_{ion}$ is the ionizing luminosity illuminating the wind, $n$ is the density of the wind, $R_{sph}$ is the distance of the wind from the ionizing source, discussed in detail in section \ref{sec_xstar}.} (see discussion next section). The spatial oscillations seen in the disk region (within red zone) are characteristic non-linear `channel mode' behaviors due to the magneto-rotational instability (see discussion section 3.1 in \citealt{Jacquemin2019}). As $\mu$ increases, the disk displays from 3 (left) to only 1 (right) oscillations, highly magnetized (near equipartition) solutions having no such oscillations.       

Since all solutions used here share the same ejection efficiency $p\sim 0.1$, the jet asymptotic speed is nearly the same. However, the wind geometry and therefore both velocity and density profiles are actually changing with $\mu$ in the acceleration zone near the disk. This is best seen in Fig.~\ref{fig_mag_field_geometry} where, for each solution, we show only one magnetic surface anchored at $6 R_G$ ($R_G$ is the gravitational radius). The black solid line is our chosen line of sight (LOS) with an inclination of $i=21^o$. It can be seen that, as $\mu$ increases, the magnetic surfaces get more and more vertical (all solutions eventually recollimate toward the axis, \citealt{Ferreira1997, Jannaud2023}). This is because the magnetic tension due to the poloidal field increases with $\mu$. Note that this effect has nothing to do with the hoop-stress, which is related to the toroidal field and plays a role only beyond the Alfv\'en surface. The poloidal magnetic tension, which acts against the inertia of the outflowing plasma and tends to close the magnetic surface, is mostly effective in the sub-Alfv\'enic region \citep{Ferreira1997}. 
Thus along a given LOS, for higher magnetization solutions, the wind solutions start at slightly closer (see the inset) distances from the black hole.

The density and the velocity distributions of the wind play crucial roles in deciding the absorption line strength and profile respectively. The variation of density and velocity along the LOS is shown in Fig.\ref{fig_vel_den_pro}. Although the disk magnetization $\mu$ affects both the density and velocity, the variations in density (Fig.\ref{fig_vel_den_pro}, panel B) are not as significant as those for the velocity profile (Fig.\ref{fig_vel_den_pro}, panel A). 

It can be seen that, as $\mu$ increases, the projected velocity significantly decreases (factor 4). This projection effects leads therefore to this somewhat counter intuitive result that, the smaller the magnetic field, the larger the detected wind velocity. This is something to keep in mind when interpreting line profiles.  

Note that in Fig.\ref{fig_vel_den_pro}, the range of radial distance from the black hole (x-axis) extends outwards from $\sim 500 \, R_g$. Of course, according to our model formulation, outflowing material exists in the region inwards of $\sim 500 \, R_g$ as well. However the range emphasised in the figure is for the part of the outflow which has low enough ionisation so that it can `significantly' absorb photons resulting in Fe XXVI (and lower ions) absorption lines in the X-ray spectrum. Discussions in the second paragraph of section \ref{sec_xstar} and Appendix \ref{sec_justification_rad_wind} detail the method of how a cutoff of log$\xi\leqslant 6$ was used to select the wind region with low enough ionisation.

\subsection{Selection of line of sight for theoretical spectra}

In the disk-wind system, optically thin wind is launched from an optically thick disk which lies at the equatorial region. Consequently, for a typical accretion-ejection solution, a LOS too close to the equatorial region becomes Compton thick (N$_H\geqslant1.5\times$10$^{24}$ cm$^{-2}$). This constrains the minimum inclination angle below which wind signatures can not be observed. 

On the other hand, if the wind is heavily photoionized, no absorption is possible. Since photoionization depends inversely on the density, in the polar regions the wind becomes easily over-ionized due to low densities as can be seen in Fig. \ref{fig_Nh_xi_distribution}. Using these two constraints (Compton thick at low $i$ and over-ionization at large $i$), one can find the detectable region of the wind (for more details, see Paper I). In this work, we assume that the disk is extended up to a distance of 10$^5 R_g$ from the black hole, fixing thereby the distance at which the last magnetic field line is anchored.

We present the spatial distribution of log($\xi$) in the r-z plane for $\mu=$ 0.0010, 0.0157, 0.0674 in Fig.  \ref{fig_Nh_xi_distribution}. The Compton thick region is shaded in red. The lowest LOS, which is Compton thin, is shown as a straight white line, and the corresponding value of $i$ is mentioned next to it. Density contours are shown in blue dashed lines, and values next to them give an understanding of how density changes with LOS. Paper II showed that variation of LOS can lead to a significant change in the absorption spectra, for MHD models whose density falls fast with increase in $i$. Here, in this paper, we solely concentrate on the effect of magnetization on the spectra and hence fixed the LOS at $i = 21^\circ$ throughout the work. The luminosity calculated in Appendix \ref{sed_details} is used here to find the ionization distribution.

\section{Incident ionising spectrum}
\label{sec_in_spectra}

\begin{figure}
\centering\includegraphics[width=\columnwidth]{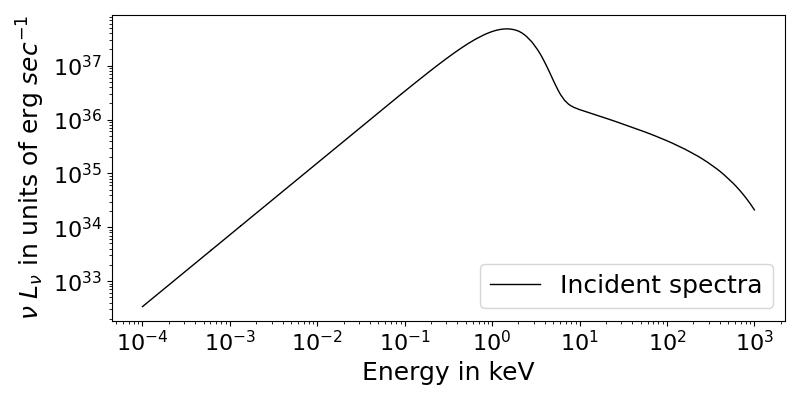}
\caption{Spectral energy distribution of the ionising radiation emitted from the innermost vicinity of the black hole, and illuminating the wind.}
\label{fig_in_spectra}
\end{figure}

The spectral energy distribution (SED), both its shape and normalization, determines how the material in the outflow will be photoionised, which in turn determines the transmitted spectra. The SEDs of black hole XRBs have two distinct components \citep{Remillard2006, Done2007}: (1) a multi-temperature blackbody component and (2) a non-thermal power-law component. To model the multi-temperature disk component, we have followed the same procedure as mentioned in \cite{Bhat2020}. The power-law component had then been added following the prescription of Paper II. The required elaborate details are also recorded in Appendix \ref{sed_details}. In this section we summarise and provide only the necessary parameters for the reader's information.

To generate the thermal multi-temperature blackbody component of the SED, we use $diskbb$ model \citep{Mitsuda1984, Makishima1986} in {\sc XSPEC} \footnote{https://heasarc.gsfc.nasa.gov/xanadu/xspec/} \citep{Arnaud1996}. A hard power-law with a high energy cut-off ($h\nu_{max} = 100$ keV) and low energy cut-off ($h\nu_{min} = 20$ eV) is then added to $diskbb$ component ($f_{disk, \nu}$) to generate a fiducial Soft state SED, 
\begin{equation}
f_{\nu}(\nu)=f_{disk, \nu}(\nu)+[A_{pl}\nu^{-\alpha}\times exp^{-\frac{\nu}{\nu_{max}}}] \times exp^{-\frac{\nu_{min}}{\nu}}.
\label{eq_input_fnu}
\end{equation}
The normalization factor of the power-law ($A_{pl}$) is adjusted such that the disk contributes 80\% of the 2-20 keV flux. The values of all the parameters involved here and the method in more details are written in Appendix \ref{sed_details}. The SED prepared above with appropriate normalization is shown in Fig. \ref{fig_in_spectra}.

The self-consistency of the MHD accretion-ejection solutions we are using is that the density of the wind is linked to the disk through the accretion rate. In this work, we impose this consistency between our flow density and the luminosity of the incident SED by assuming that the accretion rate provided in $diskbb$ (to find the luminosity), is the same as that used to find the density (and subsequently, the ionization) distribution of the wind.

\section{Computation using XSTAR}
\label{sec_xstar}
In earlier works (Paper I \& Paper II), we used photoionization code $CLOUDY$\footnote{URL: https://trac.nublado.org/} (\citealt{Ferland1998, Ferland2017}; version C08.00) to compute the transmitted spectra. However, for wind observations from XRB, Fe XXVI Ly-$\alpha$ doublet can play a crucial role (\citealt{Tomaru2020}), which cannot be accurately modeled using $CLOUDY$  (we checked up to the latest version C23.00). Fe XXVI Ly-$\alpha$ doublet will be an important part of our analysis in this paper and also all of our subsequent studies on black hole XRB spectra. Hence we choose XSTAR (version 2.54a) to synthesize the transmitted spectra. For greater flexibility, we use the subroutine version of XSTAR\footnote{https://heasarc.gsfc.nasa.gov/xstar/docs/html/node181.html}. Although we follow the usual procedure (Paper I, \citealt{Fukumura2017}, Paper II), for the sake of completeness, we describe the method in some detail in Appendix \ref{sec_step_by_step}, where the reader can also find discussion on the input parameters required by XSTAR. To do this kind of study, it is customary to handle the radiation and the wind separately which is justified in Appendix \ref{sec_justification_rad_wind}.

\begin{figure} [t]
\centering\includegraphics[width=\columnwidth]{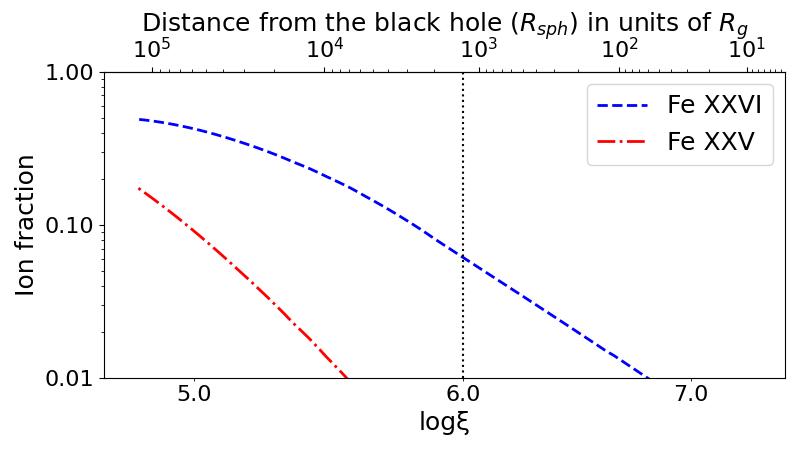}
\caption{Variation of ion fraction with the ionization parameter, log$\xi$ for the MHD model of $\mu=0.0674$ along LOS of $i = 21^{\circ}$. Ion fraction drops rapidly at higher ionization, and at log$\xi=6.0$ Fe XXVI becomes less than 10\% of Fe, whereas Fe XXV becomes almost zero. Essentially no absorption will take place for log$\xi\gtrsim6.0$, for these ions. Distance from the black hole is shown in the upper x-axis. We use log$\xi=6.0$ as a marker to start the XSTAR computation (\citealt{Chakravorty2013})}
\label{fig_Fe_fraction}
\end{figure}

To simulate the spectra from the wind which is spread across a large extent ($\sim 10^5 R_g$), we split the wind region into slabs by assuming a logarithmic radial grid $\Delta R_{sph}/R_{sph}=0.115$, where $\Delta R_{sph}$ is the width of each slab and $R_{sph}$ is the location (midpoint) of the slab. The choice of the radial resolution is explained in Appendix \ref{sec_size_of_box}. For the computation of transmitted spectra from each slab through XSTAR, we need to provide density, column density, and ionization parameter ($\xi=L_{ion}/(nR^2_{sph})$) of the slab, where $L_{ion}$ is the ionizing luminosity illuminating the wind and also the input luminosity parameter in XSTAR. We compute $L_{ion}$ by integrating the incident spectra (equation \ref{eq_input_fnu}) in the energy range 1-1000 Rydberg and assuming the source to be isotropic (equation \ref{eqn_L_ion}). The calculation of ionizing luminosity is further detailed in Appendix \ref{sed_details}. We already have the rest of the required inputs from sections \ref{sec_MHD_solns} and \ref{sec_in_spectra}. We thus have a series of slabs whose physical properties are dictated by MHD solution and the ionising radiation. Note that the slabs (and hence the outflow) is considered to be present from $6 R_g$ outwards. However, the ion fraction of Fe XXVI falls below 10\% above $\log \xi = 6$ - see Fig. \ref{fig_Fe_fraction} for the solution of $\mu=0.0674$. Hence we choose this value of $\log \xi=6$ as an upper limit for our XSTAR calculations. For each MHD solution and line of sight we find the slab for which $\log \xi = 6$ and conduct XSTAR calculations from this slab and outwards. Although $L_{ion}$ is the same throughout this work, different density distributions for the different outflow models with different $\mu$ make the ionization distribution different, and consequently, the starting radius for absorption changes. The outflowing MHD wind also has a wide velocity range (Fig. \ref{fig_vel_den_pro}(A)), which is likely to Doppler shift the spectrum seen by each slab, as well as the final spectrum seen by the observer. This effect is crucial for creating the shape of the asymmetric line profile due to MHD wind. This whole method of Doppler shifting the SEDs is similar to what was done in Paper I, II, \citealt{Fukumura2017} and the process has been discussed further in Appendix \ref{sec_dopplershift}.

\begin{figure*}
\centering\includegraphics[width=\textwidth]{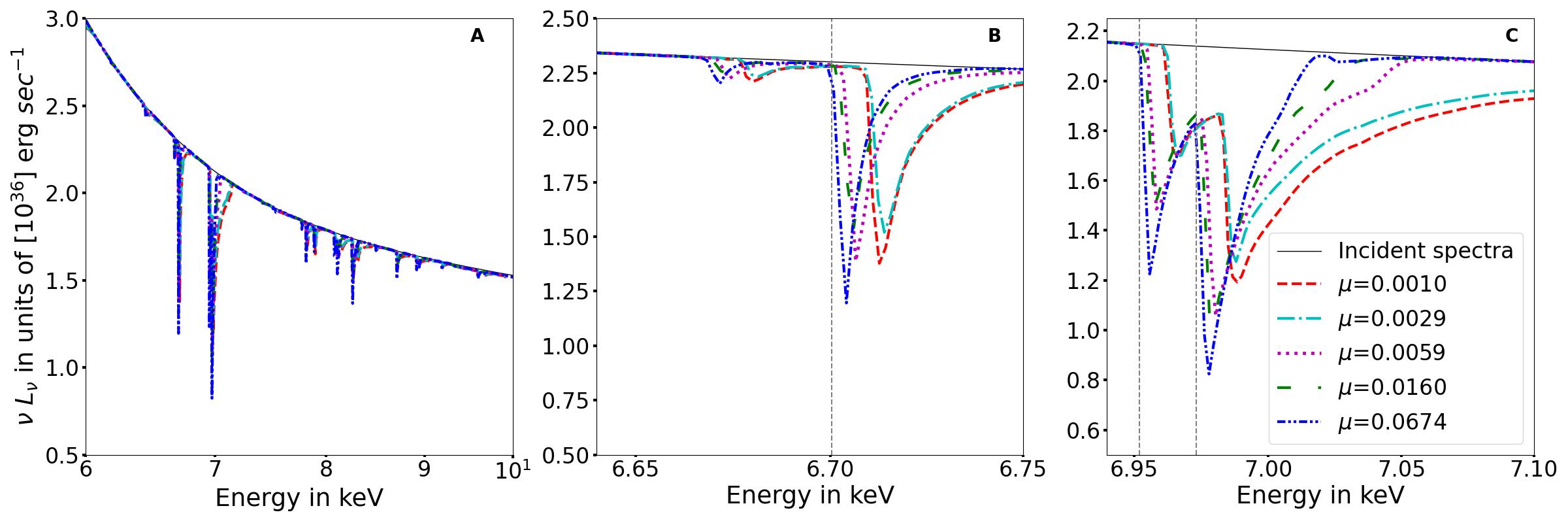}
\caption{Variation of transmitted spectra for different magnetizations. The angle of LOS is $i=21^\circ$ from the plane of the disk. The cutoff of log$\xi=6$ sets the inner boundary. The outer boundary is set by the field line anchored on the disk at a distance of 10$^5 R_g$ from the black hole. Broadband view of 6-10 keV is presented in panel (A) indicating the most prominent absorption lines are Fe XXV and Fe XXVI, zoomed in view of which are shown in panel (B) and (C) respectively. Rest frame energy of Fe XXV resonance line (6.70040 keV) in panel (B) and, Fe XXVI Ly$\alpha_1$ (6.97316 keV) and Ly$\alpha_2$ (6.95197 keV) in panel (C) are indicated by the vertical grey lines. Magnetization clearly plays a decisive role in shaping the transmitted spectra.}
\label{fig_ref_transmitted_spectra}
\end{figure*}

\begin{figure*}[b]
\centering\includegraphics[width=\textwidth]{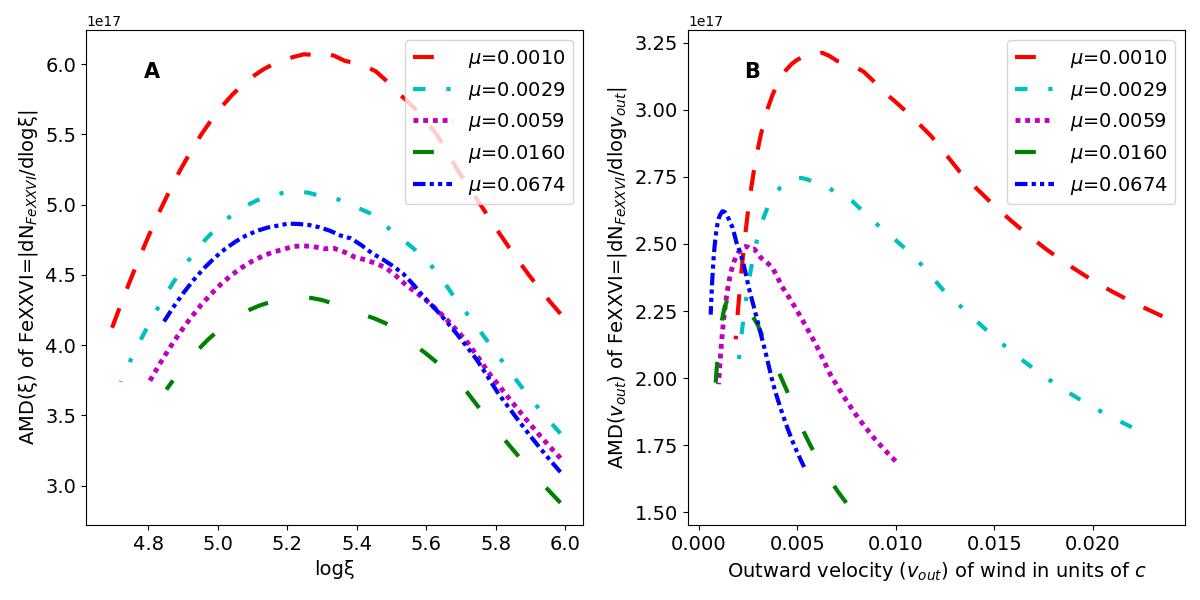}
\caption{For Fe XXVI, AMD($\xi$) versus log$\xi$ is plotted in panel A and panel B shows the AMD($v_{out}$) with respect to $v_{out}$.}
\label{fig_AMD}
\end{figure*}

\section{Results}
\label{sec_results}
In this section we take deep looks at the theoretical spectra computed using Xstar as well as the ones derived after convolving the theoretical spectra with instrument responses of XRISM and Athena.

\subsection{Model spectra and their variations with $\mu$}
\label{sec_variation_with_mag}

\subsubsection{Model spectra}

In Fig. \ref{fig_ref_transmitted_spectra}, we plot the variation in transmitted spectra for the different CLM solutions with varying disk magnetization. Panel (A) shows the 6-10 keV spectra, whereas panel (B) and panel (C) focuses on the Fe XXV and Fe XXVI line, respectively. These XSTAR output spectra are termed as model spectra to distinguish from fakespectra generated using instruments' response later. 

Fe XXVI is one of the very few ions which shows significant absorption features in this heavily ionized medium, and due to spin-orbit coupling, Fe XXVI Ly-$\alpha$ line splits into doublet with rest frame energy at 6.952 and 6.973 keV. Fig. \ref{fig_ref_transmitted_spectra}(C) focuses on the Fe XXVI Ly-$\alpha$ doublet. Panel B concentrates on the Fe XXV absorption line. In both panels we notice that the blue shift and the broadening of the line increases significantly with decreasing magnetization. This result was hinted by the evolution of wind velocity with disk magnetization (Fig. \ref{fig_vel_den_pro}(A)) in section (\ref{sec_hydro}). In this section we analyse deeper as to how the physical parameters of the outflow can influence the extent of absorption and results in the line profiles that we are seeing.

\subsubsection{Absorption Measure Distribution}

The most suitable way to understand the variations in the line profiles (as a function of $\mu$) is to study the absorption measure distribution (AMD) (\citealt{Holczer2007, Behar2009}) which is the variation of absorption with respect to the ionization parameter $\xi$. The popular way to represent the AMD is to plot the `equivalent' hydrogen column density along the line of sight as a function of log$\xi$. To focus on a specific ion, the hydrogen column density can be replaced by the ionic column density. For Fe XXVI ion, we can write
\begin{equation}
AMD(\xi)=|dN_{FeXXVI}/d(log\xi)|
\end{equation}
Here, $dN_{FeXXVI}$ represents the Fe XXVI ionic column density of each slab in our computation, and $d(\log\xi)$ is the ionization spread within the slab. The advantage of using AMD($\xi$) is that for self-similar wind solutions, AMD($\xi$), [or any other AMD(x)] becomes independent of the slab's width and scales only with $\xi$ [the physical parameter x]. For details, refer to section 3 of \citealt{Behar2009}. 

Panel (A) of Fig. \ref{fig_AMD} shows the AMD($\xi$)'s. We see that irrespective of the MHD model, maximum absorption happens from slabs which have log$\xi\sim5.3$, indicating the well-established fact that absorption is driven by ionization distribution. The upper limit of log$\xi=6.0$ where all the AMD's cut off, arises due to the choice we made (see last paragraph in section \ref{sec_xstar}). At the lower limit, different AMD's cut off at different values of $\xi$ - even if for all models we consider the disk to be $10^5 R_g$, because of the varying geometry of the flow (varying shapes of the field lines) we get different densities for different $\mu$ solutions (Fig. \ref{fig_vel_den_pro}(B)). 

The ionisation parameter however does not relate the absorption to the more direct physical parameters i.e., velocity, density of the outflow. We choose to extend the concept of AMD as a function of outward velocity ($v_{out}$), or LOS velocity of the wind, as shown in panel (B) of Fig. \ref{fig_AMD}. We choose the physical parameter velocity because, velocity reflects directly in high resolution spectra as blue-shifts of absorption lines or as the spread and asymmetry in the line profiles. We immediately see that unlike the different AMD($\xi$)'s which had more or less similar shape, the different AMD($v_{out}$)'s are diverse in shape. The peak of the ones with lower $\mu$ is at higher velocities and the AMD's are also spread over a much larger velocity range. For example, the peak of the AMD($v_{out}$) moves from $0.001c$ for the model with $\mu=0.0674$ to $0.006c$ for the one with $\mu = 0.0010$. For these respective models, the velocity range $\Delta v_{out}$ over which absorption is happening also increases from $0.005c$ to $0.02c$. The effect is directly reflected in the corresponding line spectra - most prominently seen in Fig. \ref{fig_ref_transmitted_spectra}(C) particularly via the Fe XXVI Ly$\alpha_1$ line, which is the most (least) blue shifted for the $\mu = 0.0010 \, (0.0674)$ model and has the largest (smallest) spread and asymmetry in the line profile. For an asymmetric line profile it is challenging to calculate its blue-shift. Since the peak of the AMD($v_{out}$) denotes maximum absorption, one can crudely state that the velocity of that physical region of the outflow will most significantly influence the blue shift of the absorption line. In section \ref{sec_maximum_vel}, we make an attempt to compare the maximum theoretical velocity of the wind with the maximum velocity that can be traced by fitting the asymmetric lines in the fake spectra with multiple Gaussian components.

In Fig. \ref{fig_ref_transmitted_spectra}(C) we can easily notice the two distinct lines of Fe XXVI Ly-$\alpha$ doublet, more so for the higher magnetization solutions with $\mu=0.0059, 0.0160, 0.0674$. Since absorption occurs through a much larger velocity range for lower magnetization solutions, the individual lines in the doublet suffer through more smearing out due to gradual Doppler broadening. Hence for lower magnetization solutions the distinction between the doublet features are less apparent, but still present. 

Fig. \ref{fig_ref_transmitted_spectra}A tells us that for 6-10 keV energy range, rest of the absorption lines are quite weak relative to the Fe XXV and Fe XXVI ones. We have further checked (not shown in the Figure) that all absorption lines below 6 keV (for the physical scenario considered in this paper) are also weak. Quantification of how much weaker these lines are, is beyond the scope of this paper and we shall attempt such analysis in future publication where we will be fitting actual observations. Therefore, in the rest of the paper whenever we shall present quantitative measures of absorption we shall focus only on the Fe XXV lines and the Fe XXVI Ly-$\alpha$ doublet. Below we calculate the equivalent width of the lines in the model spectra.

\begin{figure}[t]
\centering\includegraphics[width=\columnwidth]{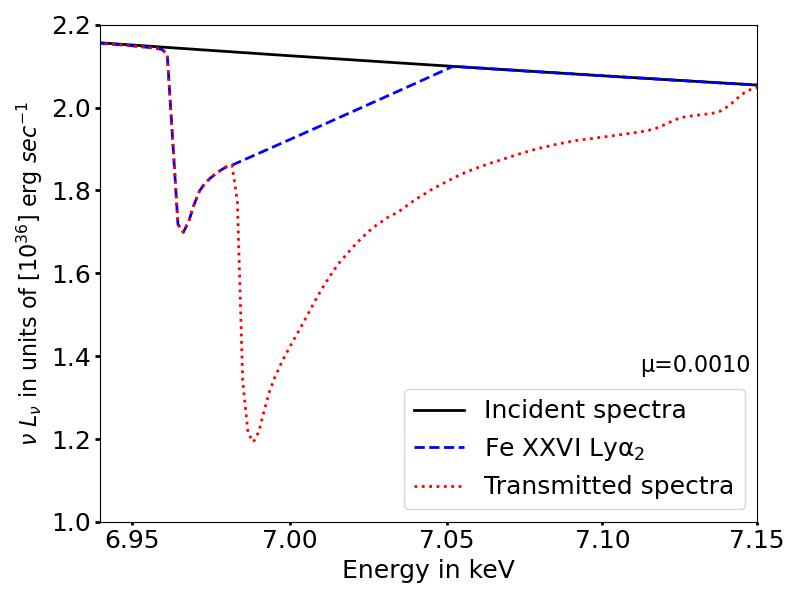}
\caption{Within the line profile of the Fe XXVI Ly$\alpha$ doublet structure (dotted, red), from the model spectrum (for the MHD model with $\mu = 0.0010$) we find the best approximation for the Fe XXVI Ly$\alpha_2$ line (in dashed blue).} 
\label{fig_eq_width_theoretical}
\end{figure}

\subsubsection{EW from model spectra}
\label{ew_variation_with_mag}

To calculate the equivalent width (EW) of any absorption line within the model spectra, we need to find the difference in the area under the incident ($L_{\nu, in}$) and transmitted ($L_{\nu, trans}$) spectra corresponding to that line. We define the EW as follows, 
\begin{equation}
\text{EW} = \int_{E_{low}}^{E_{up}} (1-L_{\nu, trans}/L_{\nu, in}) dE,
\end{equation}
$E_{low}$ and $E_{up}$ are the lower and upper bound of the energy range over which the absorption line is spread. For Fe XXV, $E_{low}=6.69$ keV, $E_{up}=6.82$ keV and for Fe XXVI, $E_{low}=6.94$ keV and $E_{up}=7.15$ keV. Following this procedure, we estimate the EW for Fe XXV and the composite Fe XXVI (consisting of the doublet) absorption lines, which are represented in Table \ref{table_equ_width}, for the various magnetizations considered in this paper. 

\begin{table*}[b]
	\centering
	\caption{Estimated EW  of Fe XXV (1s$^2$-1s2p) and Fe XXVI (1s-2p) absorption line from model spectra (section \ref{ew_variation_with_mag}) and by fitting fakespectra with multiple Gaussians (section \ref{ew_fake}) for different magnetizations. All the EWs are presented in eV. The range in brackets represents the 90\% confidence range from fitting.}
	\label{table_equ_width}
	\begin{tabular}{c|ccc|ccc}
		\hline
		\multirow{3}{*}{$\mu$} & \multicolumn{3}{c|}{Fe XXV} & \multicolumn{3}{c}{Fe XXVI}\\
		\cline{2-3} \cline{4-5} \cline{6-7}
		 & \multirow{2}{*}{model spectra} & \multicolumn{2}{c|}{fakespectra} & \multirow{2}{*}{model spectra} & \multicolumn{2}{c}{fakespectra}\\
        \cline{3-4} \cline{6-7} 
        & & XRISM & Athena & & XRISM & Athena\\
		\hline
		0.0010 & 6.3 & 6.3 (4.0-8.8) & 6.7 (4.5-8.7) & 27.1 & 31.1 (22.6-39.8) & 27.2 (19.4-35.2)\\
		0.0029 & 5.4 & 4.2 (3.1-5.5) & 6.3 (4.9-7.7) & 22.2 & 20.4 (14.4-26.9) & 28.2 (16.6-41.6)\\
		0.0059 & 3.9 & 4.6 (2.7-6.5) & 4.4 (2.7-5.1) & 18.1 & 19.8 (15.7-23.9) & 18.3 (12.6-23.2) \\
		0.0160 & 3.0 & 2.8 (2.3-3.6) & 3.0 (2.2-3.9) & 16.0 & 15.5 (12.3-18.7) & 16.1 (14.2-18.0) \\
		0.0674 & 3.5 & 3.1 (2.5-3.9) & 3.5 (2.8-4.3) & 16.6 & 15.0 (12.6-17.2) & 17.6 (11.7-23.4)\\
		\hline
	\end{tabular}
\end{table*}

To distinguish between the two components within the Fe XXVI ~Ly$\alpha$ doublet (Ly$\alpha_1$ at 6.973 keV, Ly$\alpha_2$ at 6.952 keV), however, we need to make our procedure more sophisticated. Because of Doppler broadening the right wing of Ly$\alpha_2$ (due to higher velocity absorbing gas) overlaps with the Fe XXVI Ly$\alpha_1$ (due to lower velocity absorbing gas), thus blending the two lines into the resultant profile. This is true for all the MHD models that we have considered in this paper. In Fig. \ref{fig_eq_width_theoretical} we demonstrate our attempt to separate out the two lines using the following method. We extend the right wing of Fe XXVI Ly$\alpha_2$ using a slope which is the average of the gradients (of the SED) of last two energy grid points available just before the profile falls into the trough of Fe XXVI Ly$\alpha_1$. This gives us the best estimate of the line profile for Fe XXVI Ly$\alpha_2$, and following the procedure mentioned above, we measure the EW of it. By subtracting the EW of Fe XXVI Ly$\alpha_2$ from the total EW of Fe XXVI Ly$\alpha$ doublet, we eventually find the EW of Fe XXVI Ly$\alpha_1$. The estimated EW of Ly$\alpha_1$, Ly$\alpha_2$ and their ratio from model spectra are shown in Table \ref{table_eq_width_Ly1_Ly2}. 

Ideally the ratio of EW for Ly$\alpha_1$/Ly$\alpha_2$ should be 2.0 if the absorption line is in the linear regime of curve of growth. If the absorption reaches to saturation regime, the ratio becomes lesser than 2.0. This fact is reflected from the lower values of Ly$\alpha_1$/Ly$\alpha_2$ for model spectra for $\mu=0.0160$ and $0.0674$ solutions due to very narrow velocity range over which absorption occurs and reaches to saturation regime. Ly$\alpha_1$/Ly$\alpha_2$ larger than 2.0 indicates that we underestimate the EW of Ly$\alpha_2$. 

\subsection{XRISM \& Athena-like view of the spectra }
\label{sec_simulated_spectra}

In this section, we check if the instruments with superior spectral resolution onboard XRISM and Athena are sufficient to detect the absorption lines, including the resolved doublet line profiles. We further attempt multiple Gaussian fitting within the fakespectra to handle highly asymmetric and complicated line profiles.

\begin{figure}[t]
\centering\includegraphics[width=\columnwidth]{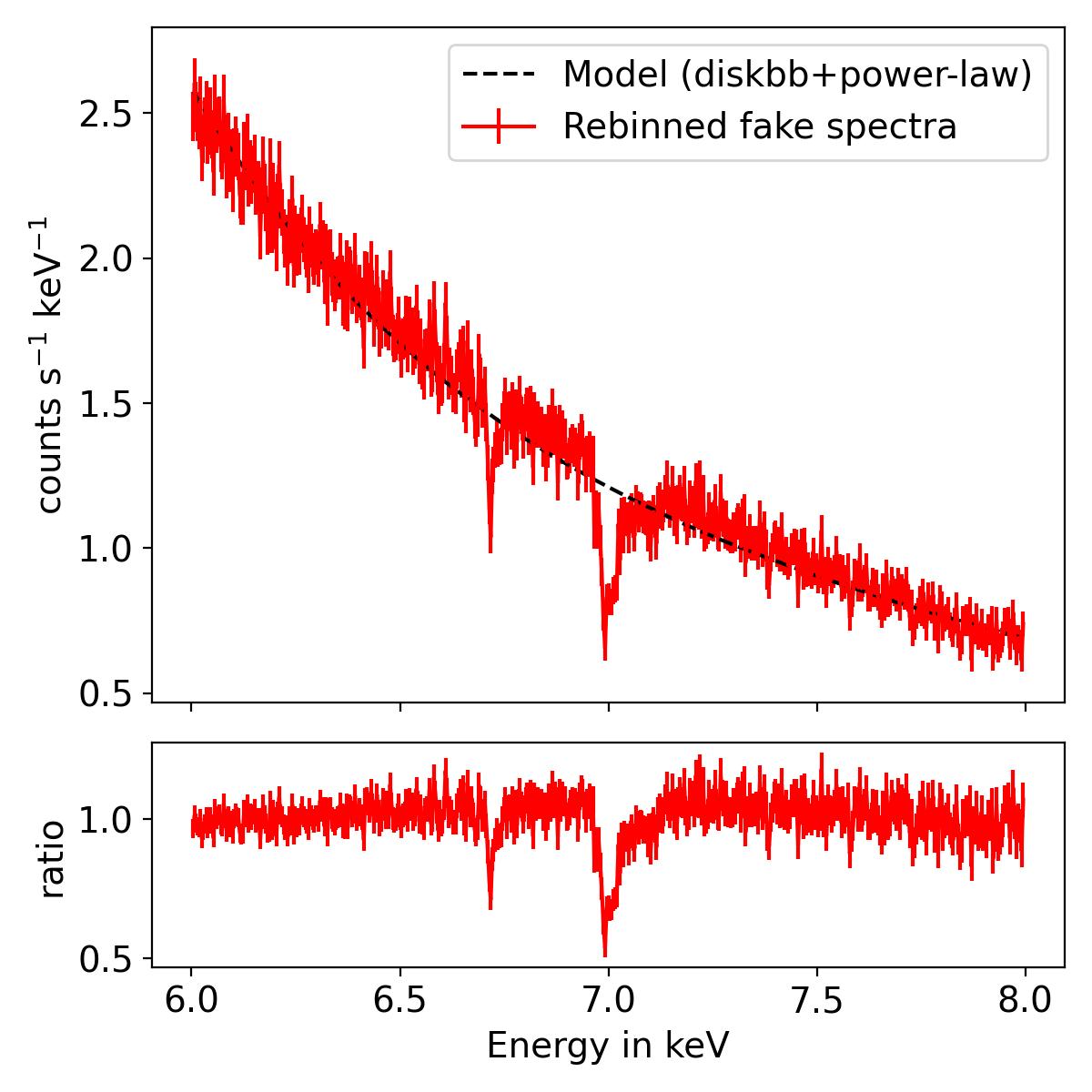}
\caption{Spectrum along i = 21$^{\circ}$ through the MHD model with $\mu = 0.0010$ convolved with XRISM response function for 100 ks exposure. The source has a flux of 3.15$\times$ 10$^{-9}$ ergs.cm$^{-2}$.sec$^{-1}$ ($\sim$ 132 mCrab) (consistent with the incident spectra generated in section \ref{sec_in_spectra}), and corresponds to a total number of 800 photon counts in the energy range 6.0-8.0 keV. \textbf{Top:} The simulated spectra (solid, red line) along with the model (diskbb+power-law; dashed, black line) to fit the continuum after optimal binning. \textbf{Bottom:} Ratio data/model.}
\label{fig_fakespectra}
\end{figure}

\begin{figure*}
\centering\includegraphics[width=\textwidth]{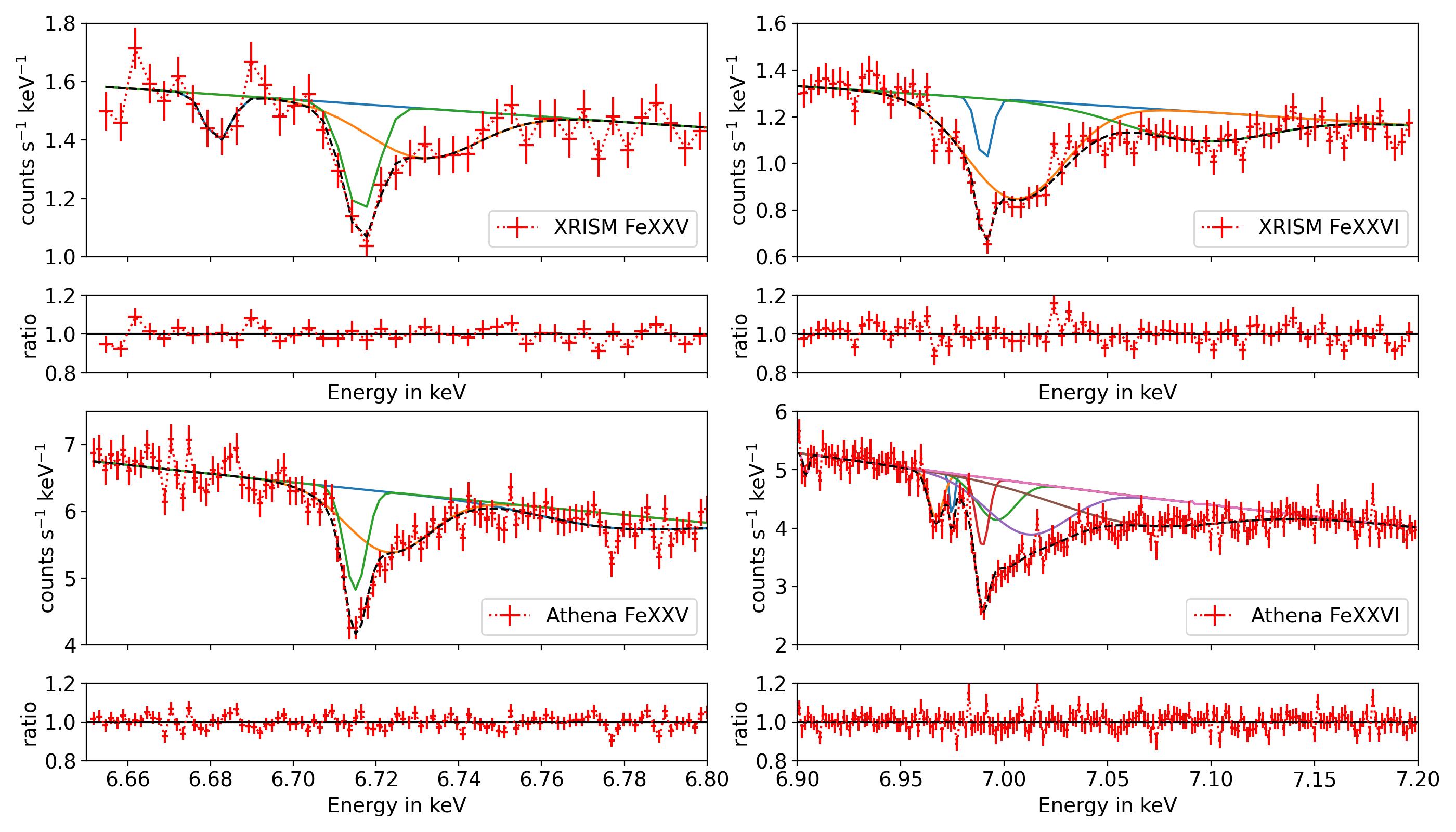}
\caption{Fitting the Fe XXV (left) and Fe XXVI (right) absorption line in the fakespectra (red, plus symbols) of XRISM (top) and Athena (bottom) with multiple Gaussians components. The fakespectra corresponds to 100 ks observations along i = 21$^{\circ}$ through the MHD model with $\mu=0.0010$. Each fit also shows the constituent Gaussian components. The total model is shown in black dashed line. The higher photon counts indicates Athena's higher effective area. The shape of the line profile is captured more accurately by Athena because of higher energy resolution. The ratio of data/model is shown at the bottom of each fit.}
\label{fig_multiple_gaussians_Fe25_Fe26}
\end{figure*}

\subsubsection{Fakespectra as proxy for future observation}

We convolve the model spectra with XRISM (\citealt{Tashiro2022}) and Athena (\citealt{Barret2023}) response for a typical standard exposure time of 100 ks using the {\sc fakeit} command in {\sc XSPEC}. Assuming the same distance to the source (8 kpc) chosen in section \ref{sec_in_spectra}, the observed flux is 3.15$\times$ 10$^{-9}$ ergs.cm$^{-2}$.sec$^{-1}$ ($\sim$ 132 mCrab) for all the $\mu$ values. After convolving the spectra with the instrument's response (see Appendix \ref{appendix_rmf_fakespectra} for the details of the response files used here and the comparison with the ones recently released during the review process of this paper. The main difference is a significant decrease of the effective area for both instruments, meaning that the exposure time used here should be significantly increased (100\% for XRISM and 40-50\% for Athena) to reach the same SNR), we always rebin the spectra using the {\sc ftgrouppha} tool in {\sc XSPEC} setting grouptype=opt, which implements an optimal binning scheme following \cite{Kaastra2016}. The convolved spectra for $\mu=0.0010$ along with the continuum model (diskbb+power-law) is shown in Fig. \ref{fig_fakespectra}, and indicates clearly that XRISM with 100 ks exposure will be able to detect the Fe XXV and Fe XXVI absorption lines. The proposed Athena telescope has an effective area $\sim$ 0.20 m$^2$ (\citealt{Bavdaz2018}) at 6 keV compared to that of XRISM which is $\sim$ 0.03 m$^2$ at 6 keV (\citealt{XRISMwhite2020}). This fact is reflected in the counts of the convolved spectra for XRISM and Athena (Fig. \ref{fig_multiple_gaussians_Fe25_Fe26}). Thus, all physical parameters remaining same, the same absorption line will be detected by Athena with much lower exposure time; or for the same exposure time, Athena can provide better statistics on line asymmetries. 

\subsubsection{EW from fakespectra}
\label{ew_fake}

To estimate EW from the observed spectra, it is typical to fit the absorption line with a Gaussian. However, for magnetically driven wind, the asymmetric absorption line shape can not be fitted well by a single Gaussian and multiple Gaussian components become necessary to fit one absorption line (method used in Paper II). The Gaussian components are added till the ftest\footnote{https://heasarc.gsfc.nasa.gov/xanadu/xspec/manual/node82.html} probability of adding another Gaussian component becomes below 90\%. Fig. \ref{fig_multiple_gaussians_Fe25_Fe26} shows the use of this method, for $\mu=0.0010$, to fit the Fe XXV (left panels) and Fe XXVI (right panels) absorption lines. We then add the EWs of the constituent Gaussians to get the effective EW of the absorption line. Results are presented in Table \ref{table_equ_width} along with 90\% confidence range (within brackets). Estimated EWs from model spectra for all the cases lie within the range of 90\% confidence limit. This implies that even if the lines are asymmetric and have very complicated profiles, the multiple Gaussian fitting methods can correctly measure of the actual strength of absorption and be proxy of the physics involved. If rigorous quantification of the individual lines of the Fe XXVI Ly-$\alpha$ doublet complex is required, then one would need even better fitting methods. While it is beyond the main scope of the paper, we briefly attempt a demonstration in Section \ref{discn_doublet}.

\section{Discussion}
\label{discn}

\subsection{MHD vs MHD-thermal outflow models}
\label{discn_sec_warm_sol}
 
In Papers I and II we found that only WHM solutions (which included `external' heating at the disk surface) were likely to yield blue-shifted absorption lines, compatible with those observed in high resolution spectra of XRBs. Further, as stated earlier as well, all those solutions (even the cold ones considered in those papers) had high disk magnetization $\mu$. One can think of the WHM solutions as MHD-thermal outflow models - which were found to be `effective' whilst the cold solutions (which were pure MHD solutions) were not dense enough. However, we see in this paper that, since the variety of pure MHD cold solutions have been enhanced by introduction of the CLM solutions, these new solutions are dense enough to produce detectable absorption lines. In this section we attempt comparison between these two kinds of solutions - WHM and CLM solutions - to understand better the relevant (for observables like absorption lines in XRBs) parameter space of MHD solutions. 

\begin{figure}[!b]
\centering\includegraphics[width=\columnwidth]{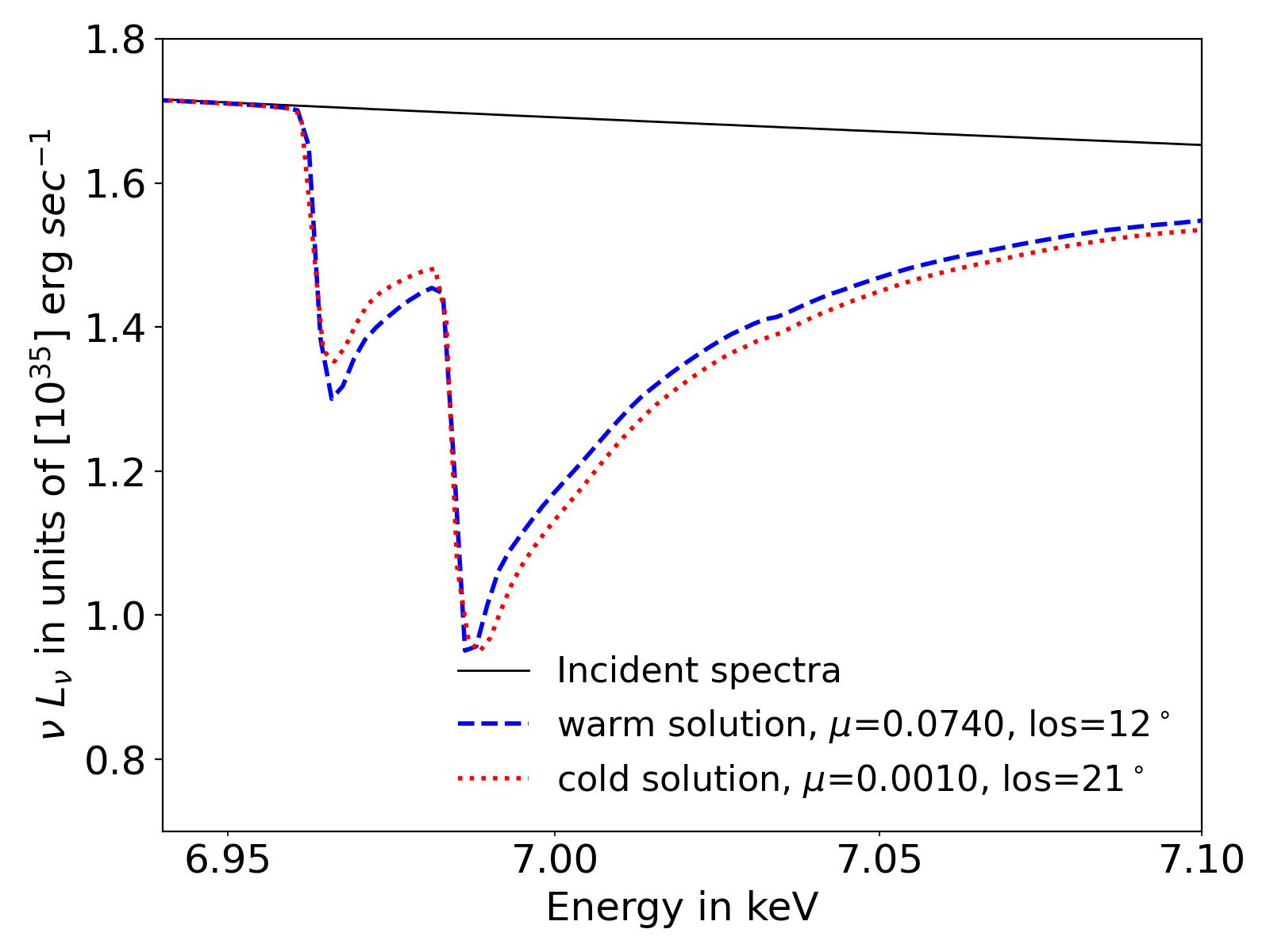}
\caption{Comparison between transmitted spectra for the WHM ($\mu=0.0740$) solution and the CLM ($\mu=0.0010$) solution. Note that to keep the density profile similar, a different LOS (closer to the disk) has been chosen for the WHM solution. Both solutions are illuminated with the same SED and the disk has the same accretion rate.}
\label{fig_comparison_warm_solution}
\end{figure}

We choose the WHM solution which has closest values for $p (= 0.1)$ and $\epsilon (= 0.07)$ compared to that of the CLM solutions [$p \sim 0.1$ and $\epsilon=0.1$] that have been used in the previous sections of this paper\footnote{The WHM solutions used in Paper II had $\epsilon=0.01$ and MHD parameters $\alpha_m=2, {\cal P}_m=1$, while values used here are $\alpha_m={\cal P}_m=1$ (see Appendix A for more details).}. According to Paper I, the slight difference in the disk thickness ($\epsilon$) should not introduce any significant change in the wind signatures. The WHM solution has a disk magnetization of $\mu=0.074$, which is at the lower end of near-equipartition solutions. Nevertheless, by comparing it with our CLM solution with $\mu=10^{-3}$, it provides a stark contrast in disk magnetization (as intended). Both the outflow models are subject to the same illuminating SED (hence the same disk accretion rate). However we allow the inclination angle to vary. While the  CLMs use a LOS i=21$^\circ$, we choose a LOS i=12$^\circ$ for the WHM because our analysis showed that although these solutions are different, it was these LOS that provided similar LOS density profile. This is because the larger the $\mu$ the larger is the accretion speed. Hence, for the same disk accretion rate, the CLM displays a denser disk, hence a denser wind. 

\begin{table*}
	\centering
    \caption{EW (in eV) of Fe XXVI Ly$\alpha_1$ and Fe XXVI Ly$\alpha_2$ estimated for different magnetizations from model spectra (\ref{ew_variation_with_mag}) and by fitting fakespectra with multiple narrow width restricted Gaussians (\ref{discn_doublet}). For the fakespectra listings, 90\% confidence range is presented in the brackets. The ratio of the two lines is presented in last two columns. }
	\label{table_eq_width_Ly1_Ly2}
	\begin{tabular}{c|cc|cc|cc}
	\hline
	\multirow{2}{*}{$\mu$} & \multicolumn{2}{c|}{Ly$\alpha_1$} & \multicolumn{2}{c|}{Ly$\alpha_2$} &
	\multicolumn{2}{c}{Ly$\alpha_1$/Ly$\alpha_2$} \\
	\cline{2-7} & model spectra & fakespectra & model spectra & fakespectra & model spectra & fakespectra\\
	\hline
	0.0010 & 19.5 & 22.5 (14.1-32.3) & 7.7 & 2.6 (1.7-3.8) & 2.5 & 8.7\\
	0.0029 & 16.6 & 21.1 (13.3-29.3) & 5.6 & 3.1 (1.5-4.6) & 3.0 & 6.8\\
	0.0059 & 12.1 & 14.3 (9.3-18.9) & 6.0 & 3.8 (2.2-4.9) & 2.0 & 3.8\\
	0.0160 & 9.8 & 12.3 (9.0-15.1) & 6.1 & 3.5 (3.1-3.9) & 1.6 & 3.5\\
	0.0674 & 9.9 & 13.3 (5.3-21.4) & 6.7 & 4.4 (3.2-5.7) & 1.5 & 3.0\\
	\hline
	\end{tabular}
\end{table*}

The observable of concern for us is the absorption line(s). We show the comparison in Fig. \ref{fig_comparison_warm_solution} - it will be impossible to distinguish between these two models in a real observation. While the existence of an external heating due to irradiation seems reasonable, and thereby, the existence of magneto-thermal winds from the outer regions of accretion disks, {\bf the same wind signatures can be accounted for by pure MHD (CLM) winds.} Here, obviously, crops up a degeneracy between the plausible models, particularly if the inclination is unknown or poorly constrained. Addressing such degeneracy is beyond the scope of this paper and we shall attempt such exercises in later publications where we shall fit actual observations.   

\subsection{Ratio of Fe XXVI Ly-$\alpha$ from fakespectra}
\label{discn_doublet}

\begin{figure}[!b]
\centering\includegraphics[width=\columnwidth]{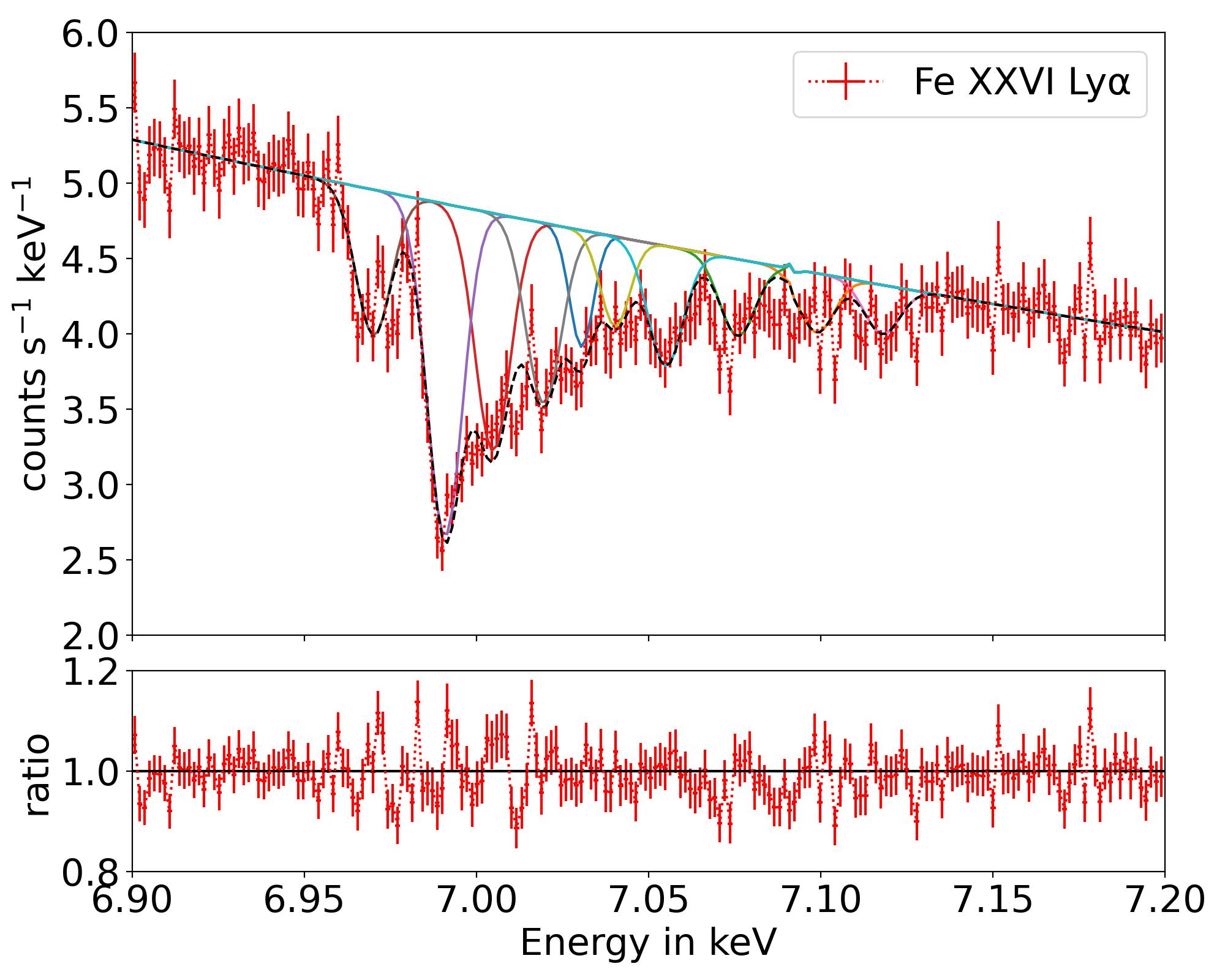}
\caption{Fitting the Fe XXVI Ly$\alpha$ absorption line in the fakespectra for Athena with multiple Gaussians while the hard limit of the line width of each Gaussian component is fixed to 5 eV. The fakespectra is computed for $\mu=0.0010$ assuming 100 ks observation of Athena (same as bottom right panel of Fig. \ref{fig_multiple_gaussians_Fe25_Fe26}). The total model is shown in black dashed line while the different Gaussian components are in different colors (solid lines).}
\label{fig_gaussians_constraining_width}
\end{figure}

The right panels of Fig. \ref{fig_multiple_gaussians_Fe25_Fe26} seem to indicate that Fe XXVI doublet line profile needs finer fitting methods than that used in Section \ref{ew_fake}. It is difficult to separate out some of the component Gaussians and assign to either Ly$\alpha_1$ or Ly$\alpha_2$ -  these `problematic' components have significant overlap in the energy ranges of both the lines of the doublet. Here we show another finer attempt with the aim to get some measure of the separate EW of the two lines of the doublet.

We restrict the hard limit for the width of the constituent Gaussians to 5 eV. Further we apply this method on the higher resolution Athena fakespectra only, where the two lines of the doublet complex is easier to distinguish than in the XRISM fakespectra. The fit for $\mu=0.0010$ (same fakespectra as in bottom right panel of Fig. \ref{fig_multiple_gaussians_Fe25_Fe26}) is presented in Fig. \ref{fig_gaussians_constraining_width}. Fe XXVI Ly$\alpha_2$ (lower energy component of the doublet) is always fitted by one or two narrow Gaussians, and no Gaussian overlaps between Ly$\alpha_1$ and Ly$\alpha_2$. Hence, simple addition of EWs of the constituent Gaussians can now give us estimates of separate EWs of Ly$\alpha_1$ and Ly$\alpha_2$. For all the $\mu$ values, the EWs and their 90\% confidence range, estimated through this procedure are given in Table \ref{table_eq_width_Ly1_Ly2}. 

Note that in this method we loose information of any large velocity tail of the Ly$\alpha_2$ that may exist within the energy range of the Ly$\alpha_1$ line. Thus this method suffer from the risk of underestimating the EW of Ly$\alpha_2$ and overestimating the EW of Ly$\alpha_1$. The line ratios are given in the last two columns of Table \ref{table_eq_width_Ly1_Ly2}. Note that the line ratios in this method are far higher than what was calculated using the previous `theoretical' method using the model spectra (section \ref{ew_variation_with_mag}). The difference between the line ratios predicted from model spectra and Athena fakespectra decreases gradually with increasing magnetization. This is not a surprising trend because with increasing magnetization the lines become narrower (less velocity spread, Fig. \ref{fig_ref_transmitted_spectra}C) and hence contamination from Ly$\alpha_2$ inside the energy range of Ly$\alpha_1$ is less. This is a general problem to estimate the EW of any absorption line whenever it is polluted by some other line. The methods that we have described may be made more accurate using the velocity information available in the outflow models. However such endeavours will come at the cost of loss of generality and also is beyond the scope of this paper.

\subsection{Maximum velocity of wind from fakespectra}
\label{sec_maximum_vel}

MHD winds are usually expected to create broad asymmetric Fe XXVI absorption lines, because the wind is present and hence absorb over a wider (than thermal winds) range of velocity.  From the fits, in Section \ref{discn_doublet}, if we compare the line centre of the most energetic Gaussian component with the rest frame energy of the Fe XXVI Ly-$\alpha_1$ line (6.97316 keV), we can get an estimate of the maximum velocity at which the wind is moving. Since in this paper, we know apriori, at what maximum velocity the wind starts to absorb (slab closest to black hole along LOS), this value can be compared to the one derived from the fakespectra. Such a comparison can help us assess how well the Athena observations can reflect the real physical scenario. The result of this exercise is presented in Table \ref{table_comparison_velocity}. We find that fitting with Gaussian components whose widths are  constrained to less than 5 eV gives us very good agreement to the `real' wind - indicating that such (or similar) techniques are quite promising to model asymmetric line profiles.

\begin{table}
	\centering
    \caption{Comparison of maximum velocity of absorbing gas along LOS estimated a) directly from MHD models (column 2), and b) by fitting Fe XXVI Ly-$\alpha_1$ line in Athena fakespectra with multiple Gaussians whose widths are restricted to 5 eV (colum 3). Velocities are presented in units of $c$.}
	\label{table_comparison_velocity}
	\begin{tabular}{c|cc}
	\hline
	\multirow{2}{*}{$\mu$} & \multicolumn{2}{c}{maximum velocity of wind} \\
	\cline{2-3} & theoretical value & from fitting\\
	\hline
	0.0010 & 0.024 & 0.021 \\
    0.0029 & 0.022 & 0.020 \\
    0.0059 & 0.010 & 0.009 \\
    0.0160 & 0.007 & 0.007 \\
    0.0674 & 0.005 & 0.005 \\
	\hline
	\end{tabular}
\end{table}

\section{Concluding remarks}
\label{sec_conclusions}

In this work, we have simulated, using XSTAR, the transmitted spectra of CLM (cold low magnetized) solutions of a new kind. These purely MHD winds are cold (isothermal) outflows emitted from the outer disk regions where a CLM solution is assumed to be settled. We varied the disk magnetization $\mu$ and looked for its influence on wind signatures, if any. Besides the disk magnetization $\mu$, the five accretion-ejection solutions used in this work share the same disk aspect ratio, disk ejection efficiency $p\simeq 0.1$ and MHD turbulence properties. The key findings are summarized below. 

 {\bf 
 (1) The CLM solutions can produce significant absorption lines. This is due to the fact that, for the same disk accretion rate, weakly magnetized disks launch winds that are denser than those from highly magnetized (near-equipartition) disks. 
 With lowering magnetization, field lines become more bent and the LOS projected velocity increases, leading to a broader and more blue-shifted absorption line. Differences in disk magnetization can therefore lead to very different line profiles.
 }

(2) Upcoming instruments, e.g., XRISM and Athena, with a typical 100 ks exposure time, will be able to detect not only the absorption line but also the line asymmetries (assuming the MHD wind is produced from the entire disk surface), which is a crucial features of MHD winds.
 
(3) For an asymmetric absorption line, fitting with multiple Gaussians is a good technique - not only for estimation of equivalent width, but also to asses physical quantities like velocity. 

(4) Asymmetry of the absorption line strongly depends on the velocity range over which absorption occurs, i.e., radial extension of the wind. MHD wind can be produced only from the outer part of the disk which seems now plausible due to weakly magnetized wind. In that case, even with MHD wind, we may not see asymmetric line profile. Although this is not explicitly shown in this paper.

{\bf  
(5) The CLM accretion-ejection solutions (pure MHD) can give rise to spectra similar to those obtained for WHM solutions (thermal-MHD).} This can be obtained by allowing a slight variation of the LOS (much less than typical observational uncertainty) and probably also by playing with other model parameters (kept constant in our study). {\bf WHM winds are therefore not a necessity anymore to describe the observed wind signatures in XRBs. 
}

\begin{acknowledgements}
SRD thanks for the support from the Czech Science Foundation project GACR 21-06825X and the institutional support from RVO:6798581. POP, MP, JF, NZ and MC acknowledge financial support from the High Energy National Programme (PNHE) of the national french research agency (CNRS) and from the french spatial agency (CNES). SB and MP acknowledge support from the European Union Horizon 2020 Research and Innovation Framework Programme under grant agreement AHEAD2020 n. 871158, from PRIN MUR 2022 SEAWIND 2022Y2T94C, supported by European Union - Next Generation EU, and support from INAF LG 2023 BLOSSOM.
\end{acknowledgements}

\bibliographystyle{aa} 
\bibliography{spectra_lmw} 

\begin{appendix}

\section{MHD solutions}
\label{appendix_MHD_solns}

The dynamics of accretion-ejection structures is intricate and complex due to the coupling between the disk and wind/jet as well as the very involvement of the magnetic field \citep{Ferreira1995, Ferreira1997, Jacquemin2019}. Here briefly, we attempt to explain the difference between highly (CHM) and weakly (CLM) magnetized cold solutions, but the interested reader is referred to these papers. 

The disk is assumed to be the locus of an MHD turbulence that leads to the presence of anomalous viscosity and magnetic diffusivities. The origin of such a turbulence is probably the magneto-rotational instability (hereafter MRI), which provides a viscosity that scales with $\mu^{1/2}$ (see e.g. \citealt{Salvesen2016} and references therein). Such a scaling appears to be consistent with the values chosen for the MHD turbulence parameters that were used in our self-similar solutions, namely $\alpha_m=1$ or 2 and a turbulent magnetic Prandtl number of unity. While viscosity provides a torque allowing accretion to proceed (although the wind torque is usually dominant), magnetic diffusivity allows the plasma to cross the field lines so that a steady-state situation can be achieved. 

CHM accretion-ejection solutions, with $\mu$ ranging from 0.1 to unity, are such that MRI is marginally quenched as the MRI wavelength is of the order of the disk scale height. Angular momentum transfer from the disk to the outflowing wind by the large scale magnetic field can then also be seen as some sort of MRI process. Instead of leading to turbulence in that case, the non-linear stage of the instability is the jet launching\citep{Lesur2013}. When the magnetic field is much weaker ($\mu$ between $10^{-4}$ and 0.1), MRI leads to the formation of non-linear channel modes within the disk, that manifest as steady-state spatial oscillations of all disk quantities (that can be seen in Fig.~\ref{fig_Nh_xi_distribution}) and leads to the generalized CLM solutions. The number of oscillations depends on the MRI wavelength \citep{Jacquemin2019}. However, despite the presence of the turbulent torque, the wind still carries away a significant portion $b= 2P_{jet}/P_{acc}$ of the accretion power $P_{acc}$ released in the disk. For all solutions used in this paper, Table~\ref{tab:MHDsolutions} provides a number of physical quantities associated with them. Note that the first five solutions are isothermal, hence CLM, and emitted from a disk with a disk aspect ratio $\epsilon=0.1$ and turbulence $\alpha_m=1$. The sixth solution is a WHM solution emitted from a disk with $\epsilon=0.07$ and $\alpha_m=2$, where the total thermal energy represents only 7.5\% of the MHD power carried away. Although it would be preferable to choose solutions with the same input parameters, i.e., identical values of ejection index ($p$), similar MHD turbulence parameters and disk aspect ratio ($\epsilon$), no WHM solution with $p=0.1$ and $\epsilon>0.07$ is available. 

Some of the involved MHD parameters are more dominant than the others in determining (and hence, causing variations in) the strengths and shapes of the absorption lines. In Papers I and II we found the model spectra are largely dominated by the ejection indices $p$ (density stratification) of the MHD outflow. The LOS angle variation and the change in size of the disk also showed some effect on the final model spectra (Paper II), but their influence was found to be less than that of $p$. In this paper we are keen in finding the influence of the variation of magnetization $\mu$ and hence we choose solutions having similar $p$ values. We found that a variation in $\mu$ has indeed a significant effect in changing the absorption lines, because of its influence on the wind geometry and projected velocity.   

Moreover, it should be noted that different $p$ values are possible within a very narrow interval for $\mu$ (see Fig.~7 in \citealt{Jacquemin2019}). Thus, over the course of Paper II and this one, we realise that there may be possibilities of degeneracy between input MHD parameters, in the future, when we try to fit observed data with our models. However, it is beyond the scope of this paper to assess how strong those degeneracies will be, and/or whether it is possible to break some of the degeneracies using constraints from other analysis methods (e.g. knowing the LOS to the source from optical light dipping analysis). In addition to the parameters of the MHD model, uncertainty in the input parameters of XSTAR can also induce changes in the model spectra which is elaborately written in Appendix \ref{sec_uncertaity_xstar_params}.

\begin{table}
    \centering
    \begin{tabular}{cccccccc} \hline
    $\mu$     & $p$ & $m_s$ & $b$ & $f_{jet}$ & $\kappa$ & $\lambda$ & $\theta_A$ \\ \hline\\
       0.0010  & 0.092 & 0.11 & 0.43 & - & 3.27 & 5.04 & 67.9\\
       0.0029  & 0.103 & 0.19 & 0.43 & - & 2.21 & 4.58 & 67.5\\
       0.0059  & 0.092 & 0.27 & 0.46 & - & 1.37 & 5.13 & 64.4\\
       0.0160  & 0.097 & 0.41 & 0.48 & - & 0.83 & 4.99 & 61.1\\
       0.0674  & 0.103 & 0.69 & 0.58 & - & 0.40 & 5.01 & 46.2\\
       0.0740  & 0.100 & 0.53 & 0.77 & 0.075 & 0.68 & 5.73 & 70.7\\ \hline
    \end{tabular}
    \caption{Parameters and quantities associated with the 6 solutions used in this paper: disk magnetization $\mu$, ejection index $p$, midplane sonic Mach number $m_s$, fraction $b$ of the accretion power carried away by the two outflows, ratio $f_{jet}$ of heat deposition per unit mass in the jet to Bernoulli integral, normalized jet mass load $\kappa$, magnetic lever arm $\lambda$ \citep{Blandford1982} and colatitude $\theta_A$ of the Alfv\'en point in degrees.}
    \label{tab:MHDsolutions}
\end{table}

\section{Details of the SED modeling and ionizing luminosity}
\label{sed_details}

The spectral energy distribution (SED) of black hole XRB has two distinct components (\citealt{Remillard2006, Done2007}); (1) multi-temperature blackbody component, (2) non-thermal power-law component. During different states of the outbursts, the BHXBs show varying degrees of contribution from the components mentioned above. According to the definition of \cite{Remillard2006}, the state where the multi-temperature blackbody component dominates and contributes more than 75\% of the 2-20 keV flux is termed as Soft state, whereas in the hard state, this contribution drops below 20\%. 

To prepare the radiation component from the accretion disk, we follow the same procedure as mentioned in \cite{Bhat2020}. A standard model to estimate the flux due to the thermal multi-temperature blackbody component of the SED is $diskbb$ (\citealt{Mitsuda1984, Makishima1986}) in {\sc XSPEC}\footnote{https://heasarc.gsfc.nasa.gov/xanadu/xspec/} (\citealt{Arnaud1996}). The required input parameters for $diskbb$ to generate the observed flux are: 
\begin{equation}
T_{in}=\left(\frac{3GM\Dot{M}}{8\pi R^3_{in}\sigma}\right)^{1/4},
\end{equation} 
the temperature at the innermost radius of the disk (\citealt{Shakura1973, Pringle1981}) and 
\begin{equation}
A_{dbb}=\left(\frac{R_{in}/km}{D/(10 kpc)}\right)^2cos \theta,
\end{equation} 
the normalization due to distance and inclination. Here we assume the mass of the black hole $M=10 M_\odot$, the accretion rate $\Dot{M}=0.1 \Dot{M}_{Edd}$, the innermost radius of the disk $R_{in}=6 R_G$ ($R_G$ is the gravitational radius), the distance to the source $D=8 \rm{kpc}$, and the angle between the LOS and the normal to the plane of the disk $\theta=70^\circ$ ($i=20^\circ$). $G$ is the gravitational constant, $\sigma$ is the Stefan-Boltzmann constant. To define $\Dot{M}_{Edd}$, we need to define accretion efficiency ($\eta$) in the equation
\begin{equation}
\label{luminosity_eta}
L=\eta\Dot{M}c^2.
\end{equation}
For the model of $diskbb$, zero torque is assumed at the inner boundary of the disk, which is termed as the standard torque scenario in \cite{Zimmerman2005}. Following equation (10) of \cite{Zimmerman2005} for standard torque scenario and comparing with our equation \eqref{luminosity_eta}, we get 
\begin{equation}
\eta=\frac{3GM}{2R_{in}c^2}=\frac{3R_G}{2R_{in}}=0.25.
\end{equation}
This is the same method as mentioned in \cite{Bhat2020} to find the flux using $diskbb$. We use this value of $\eta$ to find $\Dot{M}$ in physical unit from $\Dot{M}_{Edd}$ as
\begin{equation*}
\Dot{M}_{Edd}=L_{Edd}/(\eta c^2),
\end{equation*}
where $L_{Edd}$ is the Eddington luminosity. Following the whole procedure, we find $T_{in}=0.56$ keV as well as the flux per unit frequency $f_{disk, \nu}(\nu)$ from the accretion disk as a function of frequency ($\nu$).

Following Paper II, a hard power-law with a high energy cut-off ($h\nu_{max} = 100$ keV) is added to $f_{disk, \nu}(\nu)$ to mimic the typical observed SED (equation (\ref{eq_input_fnu})). We focus only on the soft state of the accretion flow and prepare the incident SED accordingly, as outflowing winds are observed primarily in the softer state of the accretion flow (\citealt{Miller2008, Neilsen2009, Blum2010, Ponti2012}). Following \cite{Remillard2006}, the energy index $\alpha$ is set to 1.5 (i.e., photon index $\Gamma$ to 2.5), normalization factor $A_{pl}$ is adjusted to 3.11$\times$ 10$^{-9}$ at or above 2 keV, and to (3.11/1.0005)$\times$ 10$^{-9}$ below 2 keV, such that the disk contributes 80\% of the 2-20 keV flux and the rest comes from power-law. An exponential lower energy cut-off ($h \nu_{min}=20$ eV) is introduced to diminish the contribution of the power law in the lower energy regime. For more details, see Paper II.

The luminosity radiated from the central regions of the accretion disk is what ionises the wind material. To derive this ionising luminosity, we simply integrate the model SED within the 1-1000 Ryd energy range and then multiply it by $4\pi D^2$ ($D=8$ kpc being the distance between the source and observer) 
\begin{equation}
L_{ion}=4\pi D^2\times\int_{1 Ryd}^{1000 Ryd} f_{\nu}(\nu) \,d\nu, 
\label{eqn_L_ion}
\end{equation}
where $f_{\nu}(\nu)$ is given by equation (\ref{eq_input_fnu}). We find that $L_{ion}=9.87\times10^{37}$ ergs.sec$^{-1}$. The incident SED normalized appropriately with $L_{ion}$ is shown in Fig. \ref{fig_in_spectra}. The ionization parameter $\xi$ is calculated using $L_{ion}$ for each slab depending on its density and distance, and is used as input in our XSTAR calculations. Note that, in general, equation (\ref{eqn_L_ion}) should also have had a correction factor pertaining to the LOS angle of the observer. However note that the model SED was prepared for an inclination angle of $ i = 20^\circ$ and we are studying the transmitted spectra at 21$^\circ$ thus rendering this correction factor redundant. 

\section{Details of XSTAR computation}
\label{appendix_xstar_comp}
\subsection{Inputs to XSTAR} 
\label{sec_step_by_step}
For the requirement of the photoionized cloud to be in thermal equilibrium, XSTAR\footnote{https://heasarc.gsfc.nasa.gov/docs/software/xstar/docs/html/xstarmanual.html} computes the temperature and all other physical quantities of interest e.g. ion fractions, by taking into account all possible heating and cooling mechanisms. A single computation of XSTAR assumes the photoionized cloud to have constant density. However along the LOS through the wind, the density and velocity decreases with increasing distance from the black hole following the profiles prescribed by the given MHD solution. Anchoring radius of magnetic field lines for launching wind spans a wide distance range from 6 $R_G$ to $\sim$ 10$^5$-10$^6 R_G$. Therefore, we have to split the whole wind region into slabs by assuming a logarithmic radial grid $\Delta R_{sph}/R_{sph}=0.115$, so that each slab has `near constant' density. Justification of the chosen radial resolution is discussed in Appendix \ref{sec_size_of_box}. Here $\Delta R_{sph}$ is the width of each slab and $R_{sph}$ is the location (mid point) of the slab in radial coordinate assuming the disk in the equatorial plane at $\theta=90^\circ$ ($i=0^\circ$) and the black hole is at the center of the spherical polar coordinate. The density ($n$) of each slab is fixed to the density of the wind at that location (Fig. \ref{fig_vel_den_pro}(B)). Column density ($N_h$) of each slab is calculated depending on the density and $\Delta r$.  

The atomic physics calculations within XSTAR depend on the  flux $F_{XSTAR}(\nu)$, whose SED is obviously the same as that of the input SED, but the normalisation is decided by $L_{ion}$ (eqn. \ref{eqn_L_ion}) provided as an input to XSTAR. To calculate flux from luminosity, XSTAR always assumes a spherical geometry of the cloud around the source leading to the dilution of flux from luminosity by $\frac{1}{4\pi R^2_{sph}}$, where $R_{sph}$ is the distance of the slab from the central source. 

The abundance of the wind (and hence each slab) is set to solar abundance. We always fix the turbulent velocity parameter in XSTAR to be zero throughout the whole study in this work. Resolution of energy grid is set to quite high value of 63599 for all the calculations to keep the theoretical resolution finer than the energy resolution of upcoming instruments onboard, XRISM and Athena. Other required parameters for XSTAR, i.e., minimum electron abundance, threshold ion abundance etc. are kept at default values. 

With the specified radial and energy resolution, computation of final transmitted spectra for one MHD solution takes run time $\sim$ 3-4 hours. The run time of XSTAR is almost proportional to the resolution of energy grid.

\subsection{Doppler shifting the spectrum}
\label{sec_dopplershift}

Outward velocities of the slabs are set according to the velocity profile of the wind (Fig. \ref{fig_vel_den_pro}(A)). This bulk velocity of wind Doppler shifts the spectra along LOS. The first slab is moving away from the central region of the disk. Therefore, the SED prepared in section (\ref{sec_in_spectra}) is red-shifted using the velocity of the first slab (i.e., the one with log$\xi$=6). For rest of the slabs, transmitted spectra from one slab is fed as the incident spectra of the next slab after Doppler shifting appropriately depending on the velocity difference between the two slabs. Note that the output spectrum from each slab carries the signatures of the absorption caused by the various ions in the slab. The spectra from the last slab is blue-shifted according to its outward velocity to get the final observed spectrum. For more details about the procedure, see Paper I, II, \cite{Fukumura2017}. The wind velocities we are considering are also too small to produce any relativistic Doppler correction in flux (\citealt{Luminari2020}). 

\subsection{Decoupled radiation zone and wind zone of accretion disk}
\label{sec_justification_rad_wind}

To produce the final transmitted spectrum, we assume that the radiation generated from the accretion disk passes through the wind, which is also launched from the disk. This physical picture remains consistent with our method of calculation only because we can ignore (i) the inner region of the disk as a source of radiation and (ii) the outer region of the wind as a significant contributor to absorption. 

The radiation emitted by the accretion disk (following the Shakura-Sunyaev temperature profile) within a distance of $\sim 100 R_g$ from the black hole is almost equal to the total radiation emitted by the disk. The compact size of the hot corona emitting hard X-ray power-law is also in line with this \citep{Marcel2019}. 

On the other hand, the wind near the black hole is heavily ionized due to its close proximity to the source of most energetic X-rays, thus rendering the contribution from this part of the wind as insignificant in the absorption spectrum. To proffer a quantitative idea of the absorption, we show the Fe XXVI and Fe XXV ion fraction as a function of ionization in Fig. \ref{fig_Fe_fraction} for the MHD model with $\mu=0.0674$ along a LOS of $i = 21^{\circ}$. Ion fraction of Fe XXV becomes almost zero, and Fe XXVI falls much below 0.1 in the inner region ($\sim 500 R_g$) of the wind indicating a negligible contribution in absorption (also see \citealt{Chakravorty2013}). Although, MHD outflow is present from the black hole horizon (according to our model calculations), still, upto distances $\sim$ 500-1000 $R_g$ from the black hole, the gas will not contribute to absorption. Absence of Fe XXV and Fe XXVI due to overionization confirms the absence of other `lower' ions (whose ionisation potentials are at lower energy compared to that for Fe XXV and Fe XXVI) as well. Note that we have kept the `higher ions' (e.g. He-like and H-like Neon ions) out of discussion here, because they are beyond the scope of this paper.

\subsection{Slab size}
\label{sec_size_of_box}

The MHD wind covers a large range of velocity and density, which requires splitting the line of sight into multiple individual zones that can be approximated by slabs. A higher number of slabs leads to more precise computation but will also be significantly more computationally expensive. We thus use the energy resolution of the telescopes set as reference which we should follow to simulate the spectra optimally. The wind velocity plays a major role in making the absorption line asymmetric, which is one of the key properties of MHD-driven wind. That is why in Paper I and II, wind along the LOS is divided depending on the criterion of velocity difference (or resolution) between two consecutive slabs which was set to 75 km/s at $\sim$ 6.5 keV (to keep it finer than the expected resolution of XRISM (300 km/s) and Athena (150 km/s) at 6.5 keV). 

The inner region of the wind has large outflow velocity compared to outer region (Fig. \ref{fig_vel_den_pro}(A)). 
Using the criterion of velocity resolution splits the inner region of the wind in larger number of slabs compared to that at the outer region. However we know that the outer region contributes more to absorption due to lower ionization of the wind material there. Therefore, from the point of view of contribution in absorption, outer region should be populated with more slabs (in the physical model) to provide more correct prescription of the density of wind which is used by XSTAR in estimating the final spectra. The large range in the radial distance ($\sim 10^5 R_g$) also necessitates to use a logarithmic grid. All these considerations prompt us to choose a constant value of $\Delta R_{sph}/R_{sph}$ as a criterion to split the wind into slabs. Following \cite{Fukumura2017}, we assumed $\Delta R_{sph}/R_{sph}=0.115$. However, it is equally important to check if the velocity difference between two consecutive slabs remains lesser than or equal to the resolutions of XRISM and Athena ($\lesssim$ 150 km/s). For the MHD models that we use, except for the very inner region (first 4-5 slabs out of $\sim$ 45), which also contributes the least in absorption, the velocity difference between two consecutive slabs always remain below 150 km/s.

\subsection{Uncertainties in input parameters}
\label{sec_uncertaity_xstar_params}
Possible uncertainties in the MHD model are mentioned in Appendix \ref{appendix_MHD_solns}. Moreover, there are also variants within XSTAR calculation, that can induce some changes in the model spectra. In the following, we record those effects, although a deeper quantitative analysis on those is beyond the scope of this paper. 

{\bf Geometry:} Geometry of the wind is quite uncertain which dictates how much the wind covers the incident radiation, i.e., covering fraction (in XSTAR, this is denoted as $cfrac$). Covering fraction governs the escape probability of continuum radiation in inward and outward direction. If the radiation source is completely covered by the photoionized slab (i.e., $cfrac=1$) then continuum emission in the inward direction eventually reenters the slab and the whole continuum transmits in the outward direction. Similar to continuum, this fact decreases the strength of absorption lines as well because emission lines from the other parts of the cloud adds up over the transmitted absorption lines. The reduction of strength of absorption due to covering fraction is nicely shown in Figure 4 of \cite{Trueba2019}. The dependency on covering fraction is discussed in detail in Appendix F (``Recombination Continuum Emission and Escape'') of XSTAR manual\footnote{https://heasarc.gsfc.nasa.gov/docs/software/xstar/docs/html/xstarmanual.html}. It is also very natural that radiation from LOS will be lost due to Thomson scattering as it traverses through the wind, which has been included recently in XSTAR (version 2.54a and above). The loss of radiation from LOS is a function of covering fraction and similarly as earlier, there will be no loss due to scattering for a completely closed geometry. Therefore, for partially covered geometry, due to escaping of continuum in inward direction as well as due to loss of scattering, continuum along LOS will be reduced. Reduction in continuum flux will decrease the ionization of the wind and which typically will increase the absorption in our cases. Summing up, uncertainty in the geometry of the wind puts uncertainty in the absorption line profile. In past works (\citealt{King2013, Miller2015, Trueba2019}), an average value of 0.5 is assumed. In this work, we assume completely closed geometry ($cfrac=1$) to estimate the line profile, and this gives the lower limit of absorption as decreasing the covering fraction will increase the absorption.

{\bf Turbulent velocity of the gas:} For MHD wind, absorption happens over a large velocity range ($\Delta v \sim 0.02 c$) and hence the absorption lines have significant Doppler broadening even when the gas medium is considered to be entirely photoionised. However, if there is enough thermal energy in the medium, to compete with the extent of the Doppler broadening, then a considerable value of turbulent velocity can also change the absorption line profile (\citealt{Fukumura2010}).

\section{Response files of XRISM and Athena for fakespectra simulation}
\label{appendix_rmf_fakespectra}

\begin{figure}[h!]
\centering\includegraphics[width=\columnwidth]{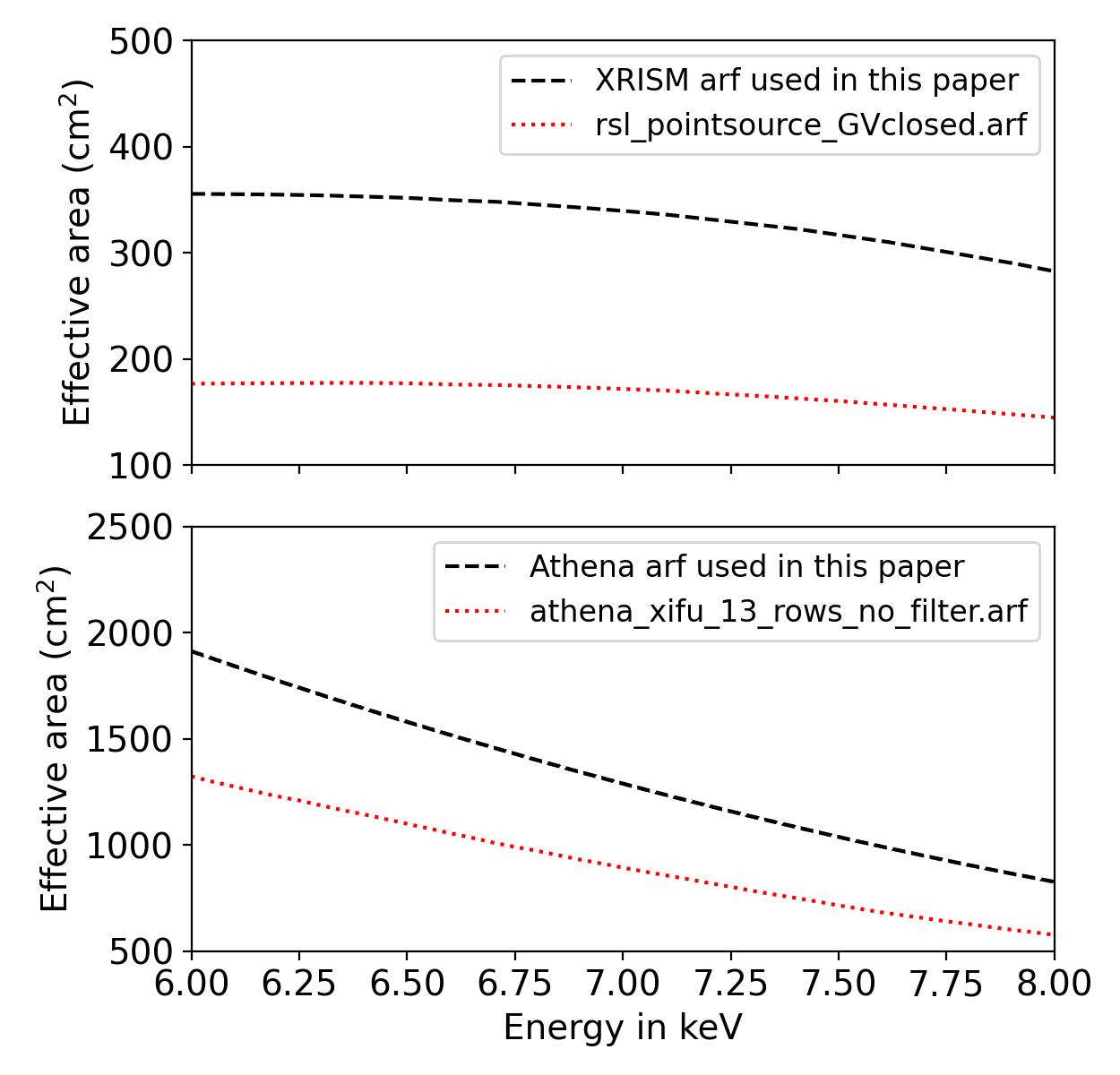}
\caption{\textbf{Top:} Comparison of effective area for XRISM between two sets of arf, one which is used in this paper to simulate fake spectra (black dashed line) and one which is available as the latest on February, 2024 (red dotted line). \textbf{Bottom:} Same as above but for Athena.}
\label{fig_eff_area_comparison}
\end{figure}

The redistribution matrix file and ancillary response file (i.e. rmf and arf) used to simulate the XRISM spectra shown in this paper are  `xarm\_res\_h7ev\_20170818.rmf' and `xarm\_res\_pnt\_pa\_20170818.arf' and correspond to the old hitomi matrices. They were the official ones before February 2024. However, the current official files are named `rsl\_Hp\_5eV.rmf' and `rsl\_pointsource\_GVclosed.arf'. They are available at \href{https://heasarc.gsfc.nasa.gov/docs/xrism/proposals/responses.html}{heasarc website of xrism}. The energy resolution is similar to the one we used, but the effective area is now very different due to the gate valve issue as shown at the top of Fig. \ref{fig_eff_area_comparison}.\\ 

For the Athena simulations, we use `XIFU\_CC\_BASELINECONF\_2018\_10\_10.rmf' as redistribution matrix file and `XIFU\_CC\_BASELINECONF\_2018\_10\_10.arf' as ancillary response file in this work. They are available on the website directory of \href{http://x-ifu-resources.irap.omp.eu/PUBLIC/OLD_RESPONSES_BEfORE_REFORMULATION/CC_CONFIGURATION/}{the old responses before reformulation} of X-IFU resources. These files have been revised recently and the official ones available in Feb. 2024 i.e. `athena\_xifu\_3eV\_gaussian.rmf' and `athena\_xifu\_13\_rows\_no\_filter.arf' are in the website directory \href{http://x-ifu-resources.irap.omp.eu/PUBLIC/NEW_ATHENA_RESPONSE_AND_BACKGROUND_FILES/}{of New Athena}. The energy resolution of this latest rmf file is 3 eV at 7 keV compared to 2.5 eV for the response used in this work. There is also a significant change in the effective area as shown at the bottom of Fig. \ref{fig_eff_area_comparison}.\\

The paper review is occurring very close to the release of the new rmf and arf files of XRISM and Athena. Therefore, we were not able to use them in the present version of the paper. Since the changes are mainly decrease of effective area, to obtain simulated spectra with similar SNR as shown in this paper, one would just have to compensate by increasing the exposure time, when using the new response files. One has to increase the exposure time by 100\% for the XRISM simulations and by 40-50\% for Athena simulations.
\end{appendix}

\end{document}